\documentclass[a4paper,11pt]{article}

\usepackage[margin=1.14in]{geometry} 
\usepackage{tabularx}
\usepackage{multirow}
\usepackage{hyperref}
\usepackage{booktabs}
\usepackage{fancybox}
\usepackage{arydshln}
\usepackage{amsmath}
\usepackage{amsthm}
\usepackage{amssymb}
\usepackage{graphicx}
\usepackage{pdfpages}
\usepackage{float}
\usepackage{bm}
\usepackage[most]{tcolorbox}
\usepackage{xcolor}
\makeatletter
\makeatletter \renewcommand{\@dotsep}{10000} \makeatother
\usepackage{wrapfig}
\usepackage[utf8]{inputenc}
\usepackage{lmodern}
\usepackage{hyperref}
\usepackage{booktabs}  

\usepackage{booktabs}
\usepackage{array}
\usepackage{tabularx}
\usepackage{listings}
\usepackage{xcolor}
\usepackage{algorithm}
\usepackage{algpseudocode}
\usepackage{graphicx}
\usepackage{subcaption}
\definecolor{darkred}{rgb}{0.6, 0, 0}
\hypersetup{
	unicode,
	colorlinks,
	breaklinks,
	urlcolor=darkred, 
        linkcolor=black, 
        citecolor=darkred,
	pdfauthor={Author One, Author Two, Author Three},
	pdfproducer={LaTeX},
	pdfcreator={pdflatex}
}
 \usepackage{listings}
 \usepackage{tcolorbox}
\tcbuselibrary{skins,breakable}
\newtcolorbox{systemprompt}{
  breakable,
  colback=gray!10,
  colframe=black,
  boxrule=0.4pt,
  left=6pt,
  right=6pt,
  top=6pt,
  bottom=6pt
}

 \usepackage{xcolor}
 \usepackage{listings}
\usepackage{xcolor}

\definecolor{promptbg}{RGB}{248,248,248}
\definecolor{promptframe}{RGB}{200,200,200}
\definecolor{codebg}{RGB}{250,250,250}
\definecolor{codegreen}{RGB}{0,128,0}
\definecolor{codepurple}{RGB}{128,0,128}
\definecolor{codeblue}{RGB}{0,0,180}
\definecolor{codegray}{RGB}{100,100,100}

\lstdefinestyle{prompt}{
    backgroundcolor=\color{promptbg},
    frame=single,
    rulecolor=\color{promptframe},
    basicstyle=\ttfamily\footnotesize,
    breaklines=true,
    breakatwhitespace=false,
    keepspaces=true,
    columns=flexible,
    xleftmargin=2em,
    xrightmargin=2em,
    framexleftmargin=1.5em,
    aboveskip=1em,
    belowskip=1em,
}

\lstdefinestyle{pythonstyle}{
    language=Python,
    backgroundcolor=\color{codebg},
    frame=single,
    rulecolor=\color{promptframe},
    basicstyle=\ttfamily\footnotesize,
    keywordstyle=\color{codeblue}\bfseries,
    commentstyle=\color{codegreen}\itshape,
    stringstyle=\color{codepurple},
    numberstyle=\tiny\color{codegray},
    breaklines=true,
    breakatwhitespace=false,
    keepspaces=true,
    columns=flexible,
    xleftmargin=2em,
    xrightmargin=2em,
    framexleftmargin=1.5em,
    aboveskip=1em,
    belowskip=1em,
    showstringspaces=false,
    tabsize=4,
    morekeywords={self, True, False, None},
}

\usepackage[T1]{fontenc} 
\usepackage{amsmath}

\newcommand{\be}{\begin{eqnarray}}
\newcommand{\ee}{\end{eqnarray}}
\def\be{\begin{equation}}
\def\ee{\end{equation}}
\def\bea{\begin{eqnarray}}
\def\eea{\end{eqnarray}}

\newcommand{\gsim}{\;\raisebox{-0.9ex}{$\textstyle\stackrel{\textstyle >}{\sim}$}\;}
\newcommand{\lsim}{\;\raisebox{-0.9ex}{$\textstyle\stackrel{\textstyle<}{\sim}$}\;}
\def\lsim{\raise0.3ex\hbox{$\;<$\kern-0.75em\raise-1.1ex\hbox{$\sim\;$}}}
\def\gsim{\raise0.3ex\hbox{$\;>$\kern-0.75em\raise-1.1ex\hbox{$\sim\;$}}}

\usepackage{graphics}
\def\coll{\texttt{CoLLM }}

\usepackage{epsfig}
\usepackage{slashed}
\usepackage[utf8]{inputenc}
\usepackage{changepage}
\usepackage{multirow}
\usepackage{pstricks}
\usepackage{dcolumn}

\lstdefinelanguage{YAML}{
  keywords={true,false,null,y,n},
  keywordstyle=\color{blue},
  basicstyle=\ttfamily\small,
  sensitive=false,
  comment=[l]{\#},
  commentstyle=\color{gray},
  stringstyle=\color{red},
  morestring=[b]',
  morestring=[b]",
  literate=
   {---}{{\color{magenta}---}}3
   {...}{{\color{magenta}...}}3
   {:}{{\color{black}:}}1
}
\usepackage[sort&compress,numbers,square]{natbib}

\theoremstyle{plain}

\theoremstyle{definition}

\usepackage{graphicx, color}

\title{\coll: AI engineering toolbox for end-to-end deep learning in collider analyses} 

\vspace{8mm}
\author{\Large{W. Esmail$^{a}$\thanks{email: waleed.esmail@ruhr-uni-bochum.de} ,  A. Hammad$^{b}$\thanks{email: hamed@post.kek.jp} and M. Nojiri$^{b,c,d}$\thanks{email: nojiri@post.kek.jp \\ Package: \href{https://github.com/AHamamd150/CoLLM.git}{CoLLM GitHub}} }}

\date{ 
{}
$^a$ Institut f\"ur Kernphysik, Universit\"at M\"unster, 
Wilhelm-Klemm-Str.\\  9, 48149 M\"unster, Germany.\\
$^b$ Theory Center, IPNS, KEK,  1-1 Oho, Tsukuba, Ibaraki 305-0801, Japan.\\
$^c$ The Graduate University of Advanced Studies (Sokendai),\\ 1-1 Oho, Tsukuba, Japan.\\
$^d$ Kavli IPMU (WPI), University of Tokyo, 5-1-5 Kashiwanoha,\\ Kashiwa, Chiba 277-8583, Japan.\\
}

\begin{document}
	\maketitle
	\vspace{4mm}
	\begin{abstract}
 \normalsize{ Recent improvements in large language models have opened new opportunities for accelerating and automating scientific workflows. In parallel, modern collider analyses are becoming increasingly complex and demand substantial programming and deep learning expertise. \coll alleviates this workload by using pretrained large language models to generate physically consistent analysis code for event selection. Additionally, it automates subsequent deep learning analyses.
To further reduce reliance on programming or deep learning experience, \coll provides a graphical user interface that allows users to perform end-to-end analyses through an interactive interface. The main motivation behind \coll is to lower the coding burden and simplify the technical complexity of collider analyses, which increasingly depend on sophisticated event selections and advanced deep learning methods.
 
}
\end{abstract}
\newpage
\noindent\rule{\textwidth}{1pt}
\tableofcontents
\noindent\rule{\textwidth}{0.2pt}
\maketitle \flushbottom
\vspace{4mm}

\section{Introduction}  
\label{sec:1}

Modern collider experiments at the Large Hadron Collider (LHC) have entered an era of unprecedented data volumes and analytical complexity.  
A typical analysis involves sequences of object reconstruction, event selection, kinematic observable computation and, increasingly, multivariate classification using deep learning methods~\cite{Larkoski:2017jix,Guest:2018yhq,Radovic:2018dip,Bourilkov:2019yoi,Feickert:2021ajf}.
Each of these stages demands expert knowledge not only of the underlying physics but also of programming language and deep learning methods.
In practice, the analysis logic conceived by a physicist must be manually translated into executable code, typically involving thousands of lines spread across parsing routines, selection functions, histogram filling, and deep learning pipelines.
This translation process is time consuming and constitutes a significant bottleneck in the physics output.
Transcription errors, such as incorrect particle identification codes or inconsistent application of kinematic cuts, can propagate silently through an analysis chain and are notoriously difficult to detect. 

Recently, large language models (LLMs) have opened new possibilities for accelerating and automating scientific workflows~\cite{brown2020language,openai2023gpt4,touvron2023llama}.
In the context of high energy physics, LLMs have been explored for a variety of tasks, including literature review~\cite{Simons:2024astro,Richmond:2025feyntune}, event classification and code generation~\cite{Fanelli:2024eic}, the automation of simulation pipelines~\cite{Ndum:2024autofluka,Gendreau-Distler:2025fsj,Plehn:2026gxv,Diefenbacher:2025zzn}, and the development of agent-based analysis frameworks~\cite{Diefenbacher:2025zzn,HWresearch:2025llm4hep}.
Broader roadmaps for integrating foundation models into physics research have also been proposed~\cite{Golling:2024abg}.
These studies collectively demonstrate that LLMs can interface with complex scientific software environments and generate syntactically correct codes.
However, the direct application of generic LLMs to full collider analysis workflows faces fundamental limitations: such models lack embedded knowledge of standard high energy physics conventions, cannot natively execute or validate the code they produce, and yield non-deterministic outputs that compromise reproducibility.

In the software engineering community, LLMs are increasingly used to assist with writing software. This has led to an approach known as ``vibe coding''~\cite{ge2025survey}, where developers describe what they want in natural language and then use the AI-generated code without carefully reviewing how it works.
While vibe coding can be helpful for quickly testing ideas, it can be risky in areas where correctness is essential.
In contrast, ``vibe engineering'' describes a more careful, disciplined use of LLMs, in which experienced developers remain fully responsible for the generated code by reviewing, testing, and validating it.
The \coll framework presented in this work follows this second approach; it uses LLMs to assist with code generation while applying physics constraints, automatic validation, and predictable execution to ensure that the resulting analysis code meets the strict requirements of collider physics.

In this paper, we introduce \coll (\href{https://github.com/AHamamd150/CoLLM.git}{CoLLM GitHub}), a framework that provides an end-to-end pipeline from plain language analysis specifications to trained deep learning classifiers for collider analyses. The main goal of \coll is to make collider analyses easier to develop by reducing the amount of coding and technical effort required. Recent collider studies often use complicated event selections and advanced deep learning methods, which normally require strong expertise not only in particle physics, but also in programming and deep learning.
\coll tackles this problem by using a large language model as a domain-aware compiler. Users describe what they want to do in plain language, and the system automatically generates complete and executable Python code for the analysis. The output of this code can then be passed directly to deep learning models, based on the user’s specification.

The framework consists of two tightly integrated components: an LLM based code generation engine, guided by a physics aware system prompt encoding the LHC Olympics (LHCO) data format~\cite{lhco2006}, kinematic conventions, and standard analysis practices; and an automated deep learning module supporting multi-layer perceptrons (MLPs), graph neural networks (GNNs) and Transformer architectures for signal to background classification.
The code generation stage incorporates deterministic decoding, structured input specifications, and an automatic error correction mechanism that iteratively repairs execution failures.
\coll supports a range of publicly available LLMs from the Qwen~\cite{qwen2.5}, Llama~\cite{touvron2023llama,llama3}, and DeepSeek~\cite{deepseek2025} families, and can operate either through local inference or cloud API access.

The remainder of this paper is organised as follows.
Section~\ref{sec:collm_framework} describes the overall architecture of the \coll framework, including the LLM  code generation module and the automated deep learning pipeline.
Section~\ref{sec:installation} provides installation instructions and a quick start guide covering both the terminal and graphical user interfaces.
Section~\ref{sec:4} details the internal design of the generation engine, the physics-aware system prompt, the automatic error correction mechanism, and the deep learning classifiers.
Section~\ref{sec:5} presents a systematic validation study using five benchmark collider processes and examines the reproducibility of the generated analyses.
Section~\ref{sec:6} concludes with a discussion of current limitations and future directions.

\section{The \coll Framework}
\label{sec:collm_framework}

The \coll  framework provides a unified methodology for collider data analysis that integrates natural language interfaces with automated deep learning techniques.
In conventional collider analyses, physics concepts must be manually translated into selection code, followed by extensive coding for deep learning architectures, training, testing and hyperparameter optimization. This procedure is time consuming and prone to transcription errors, as well as inconsistencies between the intended analysis strategy and its concrete implementation. The \coll framework addresses these limitations by introducing an end-to-end  approach in which the analysis logic is specified at a high semantic level using publicly available LLMs. Following the automatic generation of the selection code, \coll deploys advanced deep learning classifiers in a fully automated manner. The current implementation includes Multi-Layer Perceptrons (MLPs), Graph Convolutional Networks (GCNs) \cite{kipf2017semi}, Dynamic Edge Convolution (EdgeConv) \cite{wang2019dynamic}, Graph Attention Networks (GATs) \cite{velickovic2018graph}, and Transformer based models for particle cloud analysis.

\begin{figure}[!h]
\centering
\includegraphics[width=1.0\textwidth]{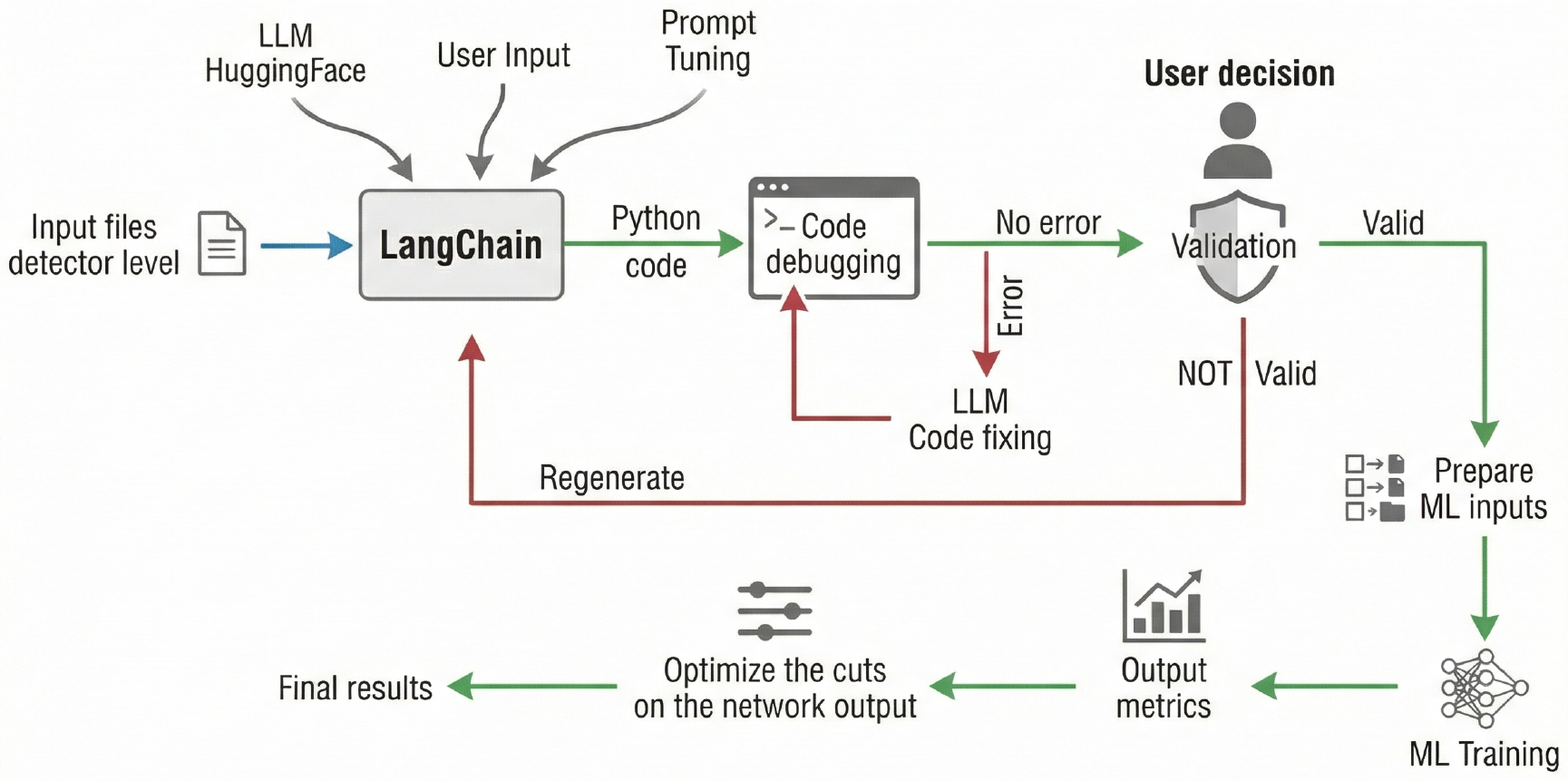}
\caption{Architecture overview of the \coll framework. The system accepts user-defined analysis specifications and LHCO data files as inputs, generates analysis code for the selection cuts, performs validation and automatic error correction, and prepares the input to the deep learning models.}
\label{fig:architecture}
\end{figure}

As shown in figure~\ref{fig:architecture}, the \coll framework consists of two coupled parts:
\begin{itemize}
\item An LLM-based module that translates plain language analysis specifications into validated event selection and preprocessing code. At present, the generated code is limited to the LHCO format, a standardized and lightweight text-based representation commonly used to store detector-level collider events.
\item An automated deep learning pipeline that consumes the output of the generated code to perform classification tasks and optimize model performance.
\end{itemize}

The sequential integration of these components ensures that downstream deep learning models are trained on features that are both physically motivated and consistently defined. This design reduces analysis time while minimizing sources of systematic ambiguity throughout the analysis chain.
\subsection{LLM for analysis code generation}
\label{subsec:llm_code_generation}

The first stage of the \coll framework uses pretrained LLMs to generate executable Python code for event preselection, kinematic feature construction, diagnostic plotting for validation, and input preparation for the selected deep learning model. This approach ensures low-level implementation details and enables the specification of analysis logic directly in natural language.
To reduce ambiguity and improve robustness, user input is organized into three semantically distinct sections:
\begin{itemize}
    \item \textbf{Selection Cuts:} specifies the definition of the signal region, including object multiplicities and kinematic or topological constraints.
    \item \textbf{Validation Plots:} defines diagnostic distributions used to verify the correctness of the implemented selections and the physical consistency of the resulting features.
    \item \textbf{Output Structure:} determines the set of observables passed to downstream deep learning models, their storage format, and the reporting of the selection summary statistics.
\end{itemize}

This structured input makes it easier for the system to understand the user's intent and helps ensure that nothing important is missed in the analysis. It also encourages users to perform basic checks and to save results in a clear, consistent way.

Code generation is performed within a closed-loop refinement procedure in which feedback from automated validation is incorporated into subsequent iterations. Validation includes static syntax checks and runtime execution tests. In the event of a failure, the model is instructed to modify only the affected code segments rather than regenerate the entire script. In practice, most failures are resolved within a small number of refinement iterations. The number of refinement cycles is defined by the user and may be disabled.

The \coll framework supports multiple pretrained LLM  available on the Hugging Face platform.  Models such as \texttt{Qwen2.5-Coder} and \texttt{Qwen3-Coder} are effective at producing syntactically correct and modular medium-sized Python code, while larger models, such as  \texttt{LLaMA-3.3-70B} and \texttt{DeepSeek-R1-Distill-Qwen-32B}, perform better in tasks with complex feature construction with large code generation. Users may select between local inference and cloud-based execution depending on the available computational resources.  LLM inference can be performed locally or via the HuggingFace cloud using a user  API key. If no API key is provided, the selected model is downloaded and executed locally. \textbf{Importantly,} if the user uses a laptop for code generation, it is recommended to use  an API key.

\subsection{Automated deep learning pipeline}
\label{subsec:ml_automation}
The second stage of the \coll framework provides an automated deep learning pipeline tailored to collider physics classification tasks. The framework supports multiple neural network architectures, each matched to a specific representation of collider events commonly used in collider analyses. A fully connected MLP is employed for fixed length, high level kinematic inputs. This model provides a simple and robust baseline for analyses based on tabular features. 

For analyses that operate directly on collections of physics objects with variable multiplicities, such as jets, tracks, or particles, GNNs are employed. In this representation, individual objects are treated as nodes and their relations as edges, allowing the model to explicitly capture object correlations while maintaining permutation invariance. This makes GNNs well suited for jet substructure studies and particle level event representations. 

Transformer architectures further extend this capability by modeling sets of particles using self-attention mechanisms. This allows the network to learn long range dependencies and global event structure without imposing a fixed relational topology. As a result, Transformers are particularly effective for particle cloud representations, where both local and global correlations are relevant.

The training procedure is fully automated while remaining configurable. Users can control dataset sizes, validation strategies, optimization algorithms, learning rate schedules and numerical precision modes, while default settings follow established practices in collider analyses. To mitigate overfitting, particularly in scenarios with limited signal statistics, early stopping, dropout, and regularization are applied by default.  Model performance is evaluated using metrics standard in collider physics, with emphasis on threshold-independent measures such as the area under the receiver operating characteristic curve (AUC). In addition to scalar performance metrics, \coll produces diagnostic plots that facilitate validation and physical interpretation of the trained classifiers, including accuracy, precision, recall  and the F1 score.

Overall, the \coll framework defines a coherent end-to-end workflow:
\begin{equation*}
\begin{split}
\text{User input in natural language}
&\;\xrightarrow{\text{LLM}}\;
\text{Executable analysis code}
\;\xrightarrow{\text{Preselection}}\;
\text{Feature dataset}\\
&\;\xrightarrow{\text{AutoML}}\;
\text{Optimized classifier}.
\end{split}
\end{equation*}
The framework is designed to maintain advances in both LLMs and collider machine learning techniques, providing a scalable and extensible tool for next generation collider analyses.

\subsection{Package structure}

The \coll framework is a modular structured package that converts plain language specifications into executable Python code for collider event analysis. In addition, it provides integrated deep learning pipelines for signal to background classification.  The main entry point of the framework is the \texttt{run.sh} script, which provides a unified command line interface for all operational modes and performs automatic dependency checks. The framework operates in five modes: (i) \texttt{TUI} mode, in which user defined plain language inputs are processed via the terminal to generate and execute analysis code; (ii) \texttt{GUI} mode, which provides an interactive web interface for configuring analyses and training models; (iii) \texttt{MLP} mode, for training multi-layer perceptron classifiers; (iv) \texttt{GNN} mode, for training graph neural network classifiers; and (v) \texttt{Transformer} mode, for training attention-based particle cloud classifiers.

The \texttt{source/} directory is divided into six main modules, each responsible for a specific part of the workflow:
\begin{itemize}
    \item \textbf{LLM Module} (\texttt{source/LLM/}):  
    This module implements the natural language interface of the framework. It includes the code generation engine (\texttt{collm\_lhco.py}), which interfaces with pretrained LLMs either locally, using the HuggingFace Transformers library, or remotely via the HuggingFace Inference API.  Local inference supports 4-bit quantization through \texttt{bitsandbytes} to reduce memory usage. The module also includes an automatic error correction component (\texttt{pyfixer.py}), which processes runtime exceptions, syntax errors and logical inconsistencies encountered during execution. Relevant error information is passed to the second  LLM to enable iterative refinement of the generated code.

    \item \textbf{DL Module} (\texttt{source/DL/}):  
    This module provides the deep learning infrastructure and implements three neural network families for collider event classification:
    \begin{itemize}
        \item \textbf{MLP} (\texttt{source/DL/MLP/}): Implements multi-layer perceptron models for kinematics distributions as inputs. It supports configurable architectures, multiple activation functions and standard training utilities, including early stopping, learning rate scheduling and mixed precision training.
        \item \textbf{GNN} (\texttt{source/DL/GNN/}): Implements graph neural network models for particle level event representations, where particles are treated as nodes and edges are weighted to unity. 
        \item \textbf{Transformer} (\texttt{source/DL/Transformer/}): Implements a particle cloud Transformer architecture with multi-head self-attention. The implementation supports configurable embedding sizes, attention heads and Transformer layers, as well as multiple pooling strategies.
    \end{itemize}
    Each submodule follows a unified structure with separate files for configuration, data loading, model definition, training and execution.

    \item \textbf{GUI Module} (\texttt{source/GUI/}):  
    This module provides an interactive graphical interface based on Streamlit application. It allows users to configure selection criteria, define model architectures and monitor execution in real time. The interface supports both LLM code generation and direct configuration of MLP, GNN and Transformer training jobs. 

    \item \textbf{Runs Module} (\texttt{source/runs/}):  
    This module contains scripts that orchestrate the execution of the \coll pipeline of the different modes. The \texttt{run\_preselection.py} script handles terminal execution, including code generation, validation and refinement. The \texttt{run\_preselection\_GUI.py} script provides equivalent functionality for GUI execution.

    \item \textbf{Configs Module} (\texttt{source/configs/}):  
    This module stores system prompts encoding collider physics conventions, LHCO format specifications and coding guidelines. These prompts guide the LLM toward physically robust code generation. The default prompt is provided in \texttt{system\_prompt.txt} and can be modified or replaced by the user for specific applications.

    \item \textbf{Utils Module} (\texttt{source/utils/}):  
    This module provides shared utility functions, including automatic dependency checking and installation, as well as parsing of YAML configuration files for terminal execution.
\end{itemize}

In addition to the \texttt{source/} directory, the framework includes other supporting directories:
\begin{itemize}
    \item \textbf{Templates} (\texttt{templates/}): Contains example configuration files for all operational modes, including templates for MLP, GNN and Transformer training.
    \item \textbf{Data} (\texttt{data/}): Provides sample LHCO and csv files for testing and validation.
    \item \textbf{Examples} (\texttt{examples user input/}): Contains example plain language inputs illustrating different analysis specifications.
\end{itemize}
This modular design of \textsc{CoLLM} allows individual components to be modified or extended independently while preserving the automated end-to-end workflow.

\section{Installation and Quick Start}
\label{sec:installation}
\coll is distributed as an open source Python package and is designed to run on a wide range of computing platforms, from personal laptops to high performance computing clusters. The framework supports Linux and macOS operating systems and requires Python version 3.9 or later. This section outlines the installation procedure and provides a brief guide for getting started.

The installation begins by downloading the source code from the public \href{https://github.com/AHamamd150/CoLLM.git}{GitHub} repository. Users should navigate to a working directory and clone the repository
\begin{lstlisting}
git clone https://github.com/AHamamd150/CoLLM.git
cd CoLLM-main
chmod +x run.sh
\end{lstlisting}

To avoid conflicts with existing software installations, the use of an isolated Python environment is recommended. The framework is compatible with Python 3.9 or newer; however, Python 3.11 is recommended to ensure compatibility with recent versions of PyTorch and HuggingFace libraries. An example using the Conda package manager is shown below
\begin{lstlisting}
conda create -n collm python=3.11
conda activate collm
\end{lstlisting}

The framework includes an automatic dependency management system and does not require manual installation. During the first execution, missing dependencies are detected and installed automatically using the \texttt{pip} command. Core dependencies include PyTorch for neural network computations, the HuggingFace \texttt{transformers} library for loading pretrained language models, \texttt{accelerate} for efficient model execution, \texttt{huggingface\_hub} for model downloading and caching, \texttt{langchain} and \texttt{langchain\_huggingface} for LLM orchestration and \texttt{streamlit} for the graphical user interface. These dependencies are checked at each execution of \coll, any missing packages are installed automatically.

Hardware acceleration plays an important role in code generation performance and determines which LLM can be executed locally. On systems equipped with NVIDIA GPUs, the framework automatically uses CUDA  and supports 4-bit model quantization through the \texttt{bitsandbytes} library, reducing memory usage and enabling the execution of large models on limited GPU resources. On Apple Silicon systems, Metal Performance Shaders (MPS) are used to accelerate inference on M-series processors. If no compatible GPU is available, the framework falls back to CPU execution, with increased inference time for large models. As an alternative, users may perform inference through the HuggingFace Inference API, which requires an API token and offloads computation to cloud resources.

\subsection{User interface}
\label{subsec:ui}
\coll provides two complementary user interfaces. The graphical user interface (GUI) offers an interactive web browser environment suited for real time monitoring. The terminal user interface (TUI) supports  fully scripted workflows, making it suitable for batch processing and large scale production analyses. Both interfaces rely on the same underlying framework, ensuring that analyses developed with the GUI can be reproduced using TUI configuration files.

\subsubsection{Terminal user interface}
\label{subsubsec:tui}

The terminal user interface provides a configuration workflow designed for batch   computing environments. All analysis parameters are specified in structured configuration files, which can be version controlled together with analysis outputs to ensure full reproducibility. The TUI uses YAML configuration files for readability and structure. A typical configuration for LLM code generation is as follow
\begin{lstlisting}[language=YAML]
Output_dir: "./output/"
DEFAULT_MODEL: "Qwen/Qwen2.5-Coder-14B-Instruct"
MAX_RETRIES: 3
Input_file: "./data/signal.lhco"
User_input: "./templates/user_input.txt"
Use_api: False
Api_key: "your_huggingface_api_key"
\end{lstlisting}

The \texttt{Output\_dir} specifies the directory where all outputs are written, including the generated Python script and histograms. The directory is created automatically and overwritten on subsequent runs. The \texttt{DEFAULT\_MODEL} defines the Hugging Face model identifier used for code generation. Model selection should reflect available hardware and performance requirements, as discussed in section \ref{subsubsec:model_selection}.
The \texttt{MAX\_RETRIES} parameter controls the automatic error correction loop. If the generated code fails to execute, the framework attempts iterative debugging and regeneration up to the specified number of retries. The \texttt{Input\_file} points to a representative LHCO file used to validate the generated code, while \texttt{User\_input} contains the plain language  specification.
Local and cloud  inference are selected via the \texttt{Use\_api} flag. Local inference downloads the model and runs it on available hardware accelerators, while API based inference offloads computation to HuggingFace infrastructure and requires an authentication token.

Execution is initiated with a single command:
\begin{lstlisting}[
basicstyle=\ttfamily\small,
frame=single,
backgroundcolor=\color{gray!10},
language=bash
]
./run.sh --run_TUI --input templates/user_input_TUI.yml
\end{lstlisting}

During execution, the framework reports progress through configuration loading, model initialization, code generation, validation, and optional error correction. Upon completion, a summary of outputs and selection statistics is printed.

In addition to analysis code generation, the terminal user interface supports configuration-driven training of deep learning classifiers. Training is fully controlled through YAML configuration files straightforward integration with batch computing systems.
Three model families are supported: MLPs, GNNs  and Transformer. Each model type is executed through a dedicated TUI mode (\texttt{--run\_MLP}, \texttt{--run\_GNN}, and \texttt{--run\_Transformer}) and uses a corresponding configuration file specifying data paths, model architecture and training parameters.
Across all model types, the configuration files define dataset splits, feature normalization, optimizer and learning rate scheduler settings, numerical precision and hardware selection. All trained models, logs, and evaluation results are written to the user defined output directory, ensuring consistent bookkeeping and reproducibility.

As an example, an MLP classification analysis can be executed using
\begin{lstlisting}[
basicstyle=\ttfamily\small,
frame=single,
backgroundcolor=\color{gray!10},
language=bash
]
./run.sh --run_MLP --input templates/config_MLP.yml
\end{lstlisting}

The corresponding YAML configuration file fully specifies the data handling, network architecture and training procedure as follows
\begin{lstlisting}[language=YAML]
seed: 42
output_dir: "output"
# ======================
# Data Configuration
# ======================
data:
  signal_path: "data/signal_mlp.csv"
  background_path: "data/bkg_mlp.csv"
  train_size: 1000  # samples per class for training
  test_size: 1000   # samples per class for testing
  val_ratio: 0.15   # fraction of training data for validation
  normalize: true   # normalize features with StandardScaler
# ======================
# Model Architecture
# ======================
model:
  layers:
    - type: dense  #dense, dropout and batchnorm
      units: 128
      activation: relu # relu, gelu, tanh, sigmoid, leaky_relu, elu
    - type: dense
      units: 64
      activation: relu
  output_units: 2  # 2 for binary classification
  output_activation: softmax  # null for logits 
# ======================
# Training Configuration
# ======================
train:
  epochs: 10
  batch_size: 256
  learning_rate: 0.001
  weight_decay: 0.0001  # L2 regularization
  optimizer: adam       # adam, adamw, sgd
  device: auto          # auto, cpu, cuda
  # Early Stopping
  early_stopping: true
  early_stopping_patience: 3  # stop if no improvement for n epochs
  early_stopping_metric: val_loss  
  # Training Precision
  precision: float32  # float32, float16, mixed
  # Learning Rate Scheduler
  scheduler:
    type: plateau  # none, step, plateau, cosine, onecycle
    patience: 5    # reduce LR if no improvement for N epochs
    min_lr: 1e-07  # minimum learning rate
  # Evaluation Metric (for model selection and reporting)
  eval_metric: auc  # accuracy, auc, f1, recall, precision
\end{lstlisting}

The \texttt{data} block specifies input datasets and dataset splitting, while optional feature normalization ensures stable training. The \texttt{model} block defines the MLP architecture as a sequence of fully connected layers with configurable activations. The \texttt{train} block controls the optimization procedure, including early stopping, learning-rate scheduling, numerical precision, and evaluation metrics. Input CSV files are adopted from earlier study in \cite{Fuks:2025qgh}

Analogous configuration files are used for GNN and Transformer training, differing only in the layer parameters. For GNNs, the configuration selects the architecture type (GCN, EdgeConv or GAT), graph construction settings and pooling strategy. For Transformers, the configuration specifies the embedding dimension, number of attention heads, number of layers and pooling method. In all cases, the training workflow remains identical, allowing direct comparison between architectures under consistent training conditions.

\subsubsection{Graphical user interface}
\label{subsubsec:gui}
The graphical user interface is implemented using \texttt{Streamlit}, a Python framework for interactive web applications. Its reactive execution ensures that interface elements update immediately in response to user input, enabling smooth interaction without page reloads. The primary role of the GUI is to construct analysis configuration files from user input and to execute the same backend scripts used by the terminal user interface. The GUI is launched via
\begin{lstlisting}[
    basicstyle=\ttfamily\small,
    frame=single,
    backgroundcolor=\color{gray!10},
    language=bash
]
./run.sh --run_GUI
\end{lstlisting}

\begin{figure}[!h]
    \centering
    \shadowbox{\includegraphics[width=0.95\linewidth]{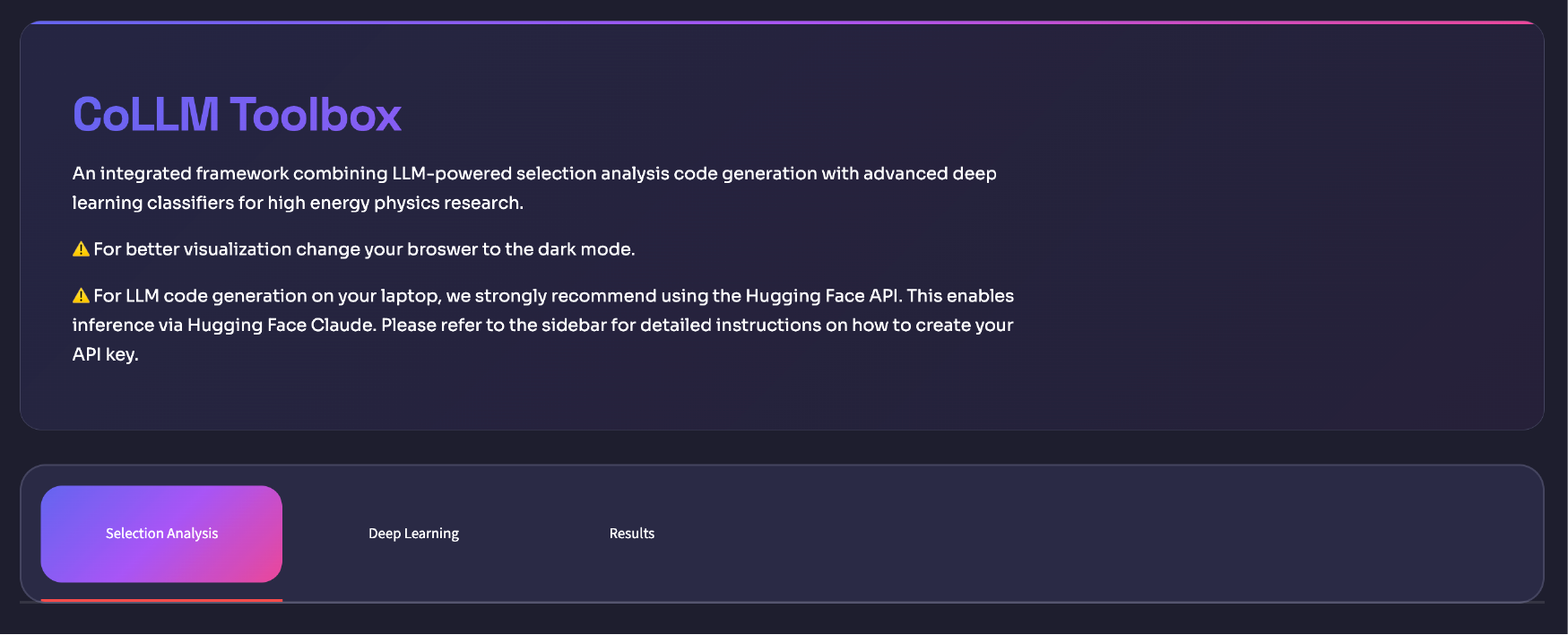}}
       \caption{Screenshot of the \coll graphical user interface.}
    \label{fig:GUI}
\end{figure}

At startup, the framework performs an automatic dependency check and installs any missing packages before initializing the Streamlit server. The interface is available at \texttt{http://localhost:8501} by default, and the port is automatically reassigned if it is already in use. The GUI provides a dark themed layout\footnote{The implemented dark theme results in improved visualization when the web browser is set to dark mode rather than light mode.} organized into three main workflow tabs, as shown in figure~\ref{fig:GUI}.
The \texttt{Selection Analysis} tab is the main interface for LLM code generation. It contains a structured text editor where users describe their analysis requirements in natural language. The editor is pre-populated with fixed section headers: \texttt{SELECTION CUTS} for object and event selection, \texttt{PLOTS FOR VALIDATION} for diagnostic histograms and \texttt{OUTPUT STRUCTURE} for cutflow reporting and feature extraction. Users must place their analysis specifications in these sections without modifying the headers. A reference panel alongside the editor provides examples of common syntax patterns, reducing the need for external documentation.

Below the editor, the LLM configuration panel controls the code generation process. Users specify the output directory for generated scripts and analysis products, as well as the path to an LHCO file used for validation. A model selection menu lists available LLMs in ascending order of capacity, ranging from lightweight models for rapid prototyping to larger models for more complex analyses. Execution can be performed either locally or via the HuggingFace Inference API; in the latter case, an authentication token is required. The maximum retries option controls the number of automatic regeneration attempts in the event of execution or syntax errors.

Pressing the \texttt{Run selection Analysis} button initiates the full generation and validation pipeline. A terminal style output panel displays real time status messages, including model loading, code generation, syntax validation, execution on test data, and any automatic correction steps. Informational messages, warnings and errors are visually distinguished to facilitate rapid diagnosis. Upon successful completion, the output location and a summary of the generated analysis are reported.

The \texttt{Deep Learning} tab configures the training pipeline that converts the extracted features into a signal to background classifier. Parameters are grouped into card-based sections for clarity. The data configuration section specifies the paths to signal and background datasets, with immediate validation of file availability. Users also define the training and test sample sizes.
The training configuration section exposes core hyperparameters, including the number of epochs and batch size, mirroring the options available in the TUI configuration files. A model selector allows users to choose between MLP, GNN  and Transformer architectures, determining the class of model to be trained.

The \texttt{Results} tab summarizes the outcomes of the training stage. Key performance metrics, including accuracy, area under the ROC curve, precision, and recall, are displayed in dedicated panels and updated dynamically when real-time monitoring is enabled. 

A persistent sidebar provides contextual assistance throughout the workflow. It includes expandable sections with guidance on obtaining and configuring a personal Hugging Face API key, without interrupting the main analysis interface.

\subsection{Example: Top quark pair production}
\label{subsec:example}

To illustrate the complete \coll workflow, we consider a representative analysis of semi-leptonic top-quark pair production. In this final state, one top quark decays leptonically ($t \to Wb \to \ell\nu b$) and the other decays hadronically ($t \to Wb \to jjb$). The resulting event topology is characterized by exactly one isolated charged lepton, missing transverse energy from the neutrino and multiple jets, including two $b$-tagged jets from the top quark decays. This example demonstrates the full analysis pipeline, from plain language specification to event selection, feature extraction and deep learning classification analysis.

The process is initiated by providing a physics specification in plain language using the structured input format stored in \texttt{example\_user\_input/user\_input\_2.txt}, together with an input LHCO file located at \texttt{data/signal\_2.lhco}. The code generation is performed using the pretrained large language model \texttt{meta-llama/Llama-3.3-70B-Instruct}. The corresponding analysis requirements are the following

\begin{lstlisting}[language=YAML]
[SELECTION_CUTS]
- Select electrons with pT > 25 GeV and |eta| < 2.5
- Select muons with pT > 25 GeV and |eta| < 2.4
- Require exactly one lepton (electron or muon)
- Require at least 4 jets with pT > 30 GeV and |eta| < 2.5
- Require at least 2 b-tagged jets
- Select MET > 20 GeV
- Select transverse mass of the lepton and MET between 30 and 150 GeV (W candidate)

[PLOTS_FOR_VALIDATION]
- Plot the following histograms:
1 Plot the transverse mass of lepton + MET (W leptonic candidate)
2 Plot the MET distribution
3 Plot the pT of the leading b-jet
4 Plot the pT of the subleading b-jet
5 Plot the invariant mass of the two leading non-b-tagged jets (W hadronic candidate) in range 40 to 120 GeV
6 Plot the invariant mass of leading b-jet + W hadronic candidate (top hadronic) in range 100 to 250 GeV
7 Plot delta R between the leading lepton and the closest b-jet
8 Plot the HT distribution (scalar sum of all jet pT)
9 Plot the number of jets per event
10 Plot eta of the leading b-jet
- Normalize all histograms to one (density=True)

[OUTPUT_STRUCTURE]
- Print summary with events before and after each cut
- Save plots in png format
- Save the following in a CSV file for MLP analysis:
  1- Transverse mass of lepton + MET
  2- Dijet invariant mass
  3- MET
\end{lstlisting}

Upon receiving these inputs, \coll processes the request using a physics aware system prompt that encodes the LHCO data format, standard four momentum reconstruction  and collider physics conventions. This prompt guides the LLM to generate analysis code that correctly parses LHCO files, handles particle type identifiers (0=photon, 1=electron, 2=muon, 3=tau, 4=jet, 6=missing transverse energy), and computes derived kinematic variables. The code generation stage produces a complete and executable Python script that includes the core LHCO parser and kinematic utility functions, shown below.

\begin{lstlisting}[language=Python]
def get_particle_mass(obj):
    """Return appropriate mass based on particle type."""
    ptype = obj['type']
    if ptype == 0:    # photon
        return 0.0
    elif ptype == 1:  # electron
        return 0.0
    elif ptype == 2:  # muon
        return 0.0
    elif ptype == 3:  # tau
        return 1.777
    elif ptype == 4:  # jet
        return obj['jmass']
    else:
        return 0.0

def four_momentum(obj):
    """Compute (E, px, py, pz) for a particle."""
    pt = obj['pt']
    eta = obj['eta']
    phi = obj['phi']
    m = get_particle_mass(obj)
    
    px = pt * math.cos(phi)
    py = pt * math.sin(phi)
    pz = pt * math.sinh(eta)
    E = math.sqrt(px**2 + py**2 + pz**2 + m**2)
    
    return E, px, py, pz

def invariant_mass(particles):
    """Compute invariant mass of a list of particles."""
    if not particles:
        return 0.0
    E_tot, px_tot, py_tot, pz_tot = 0.0, 0.0, 0.0, 0.0
    for p in particles:
        E, px, py, pz = four_momentum(p)
        E_tot += E
        px_tot += px
        py_tot += py
        pz_tot += pz
    M_sq = E_tot**2 - px_tot**2 - py_tot**2 - pz_tot**2
    return math.sqrt(M_sq) if M_sq > 0 else 0.0

def transverse_mass(visible, met):
    """Compute transverse mass of visible particle + MET."""
    dphi = delta_phi(visible['phi'], met['phi'])
    return math.sqrt(2 * visible['pt'] * met['pt'] * (1 - math.cos(dphi)))

def read_lhco(filename):
    """Parse LHCO file and return list of events."""
    events = []
    current = None
    with open(filename) as f:
        for line in f:
            line = line.strip()
            if not line or line.startswith('#'):
                continue
            parts = line.split()
            if parts[0] == '0':
                if current is not None:
                    events.append(current)
                current = {'id': int(parts[1]), 'objects': []}
            elif current is not None:
                obj = {
                    'type':   int(parts[1]),
                    'eta':    float(parts[2]),
                    'phi':    float(parts[3]),
                    'pt':     float(parts[4]),
                    'jmass':  float(parts[5]),
                    'ntrk':   float(parts[6]),
                    'btag':   float(parts[7]) if len(parts) > 7 else 0.0,
                    'had_em': float(parts[8]) if len(parts) > 8 else 0.0
                }
                obj['charge'] = 1 if obj['ntrk'] > 0 else (-1 if obj['ntrk'] < 0 else 0)
                current['objects'].append(obj)
    if current is not None:
        events.append(current)
    return events
\end{lstlisting}

The generated analysis code parses the LHCO file into structured event objects, each containing reconstructed physics objects with their kinematic properties ($\eta$, $\phi$, $p_{\mathrm{T}}$, mass, and $b$-tag). Event selection is then applied through a sequential cutflow as shown below.

\begin{lstlisting}[language=Python]
selected_events = []
for event in events:
    electrons = [e for e in get_electrons(event) 
                 if e['pt'] > 25 and abs(e['eta']) < 2.5]
    muons = [m for m in get_muons(event) 
             if m['pt'] > 25 and abs(m['eta']) < 2.4]
    leptons = electrons + muons

    if len(leptons) != 1:
        continue

    jets = [j for j in get_jets(event) 
            if j['pt'] > 30 and abs(j['eta']) < 2.5]
    bjets = [j for j in get_bjets(event) 
             if j['pt'] > 30 and abs(j['eta']) < 2.5]
    met = get_met(event)

    if len(jets) < 4 or len(bjets) < 2  or met['pt'] < 20:
        continue

    lepton = leptons[0]
    mt = transverse_mass(lepton, met)
    if mt < 30 or mt > 150:
        continue

    selected_events.append(event)
\end{lstlisting}

For the selected events, the analysis reconstructs the $t\bar{t}$ system by forming the hadronically decaying $W$ boson from the two leading non-$b$-tagged jets and combining it with the leading $b$-jet to obtain the hadronic top candidate. Several discriminating observables are then computed, including the angular separation $\Delta R$ between the lepton and the nearest $b$-jet and the scalar sum of jet transverse momenta, $H_{\mathrm{T}}$:
\begin{lstlisting}[language=Python]
for event in selected_events:
    jets = sort_by_pt(get_jets(event))
    bjets = sort_by_pt(get_bjets(event))
    met = get_met(event)
    lepton = get_leading(get_leptons(event))

    mt_values.append(transverse_mass(lepton, met))
    met_values.append(met['pt'])
    leading_bjet_pt.append(bjets[0]['pt'])
    subleading_bjet_pt.append(bjets[1]['pt'])

    non_bjets = [j for j in jets if j not in bjets]
    dijet_mass.append(invariant_mass(non_bjets[:2]))
    top_mass.append(invariant_mass([bjets[0]] + non_bjets[:2]))

    min_dr = float('inf')
    for bjet in bjets:
        dr = delta_r(lepton, bjet)
        if dr < min_dr:
            min_dr = dr
    delta_r_lepton_bjet.append(min_dr)

    ht = sum(j['pt'] for j in jets)
    ht_values.append(ht)
    njets.append(len(jets))
    leading_bjet_eta.append(bjets[0]['eta'])
\end{lstlisting}

The final step  exports the discriminating variables, after applying the selection cuts, to a CSV file  for subsequent MLP analysis:

\begin{lstlisting}[language=Python]
with open('mlp_data.csv', 'w') as f:
    f.write('mt,dijet_mass,met\n')
    for i in range(len(mt_values)):
        f.write(f'{mt_values[i]},{dijet_mass[i]},{met_values[i]}\n')
\end{lstlisting}
The exported features should  provide  kinematic information that enables the deep learning classifier to discriminate signal from background events

After code generation, \coll executes the code on the provided LHCO file and validates the output. If runtime errors occur, an automatic error correction module analyzes the traceback and iteratively  patches the code.  The validation loop continues until successful execution or the maximum number of retries is reached. 

\begin{figure}[!th]
    \includegraphics[width=\linewidth]{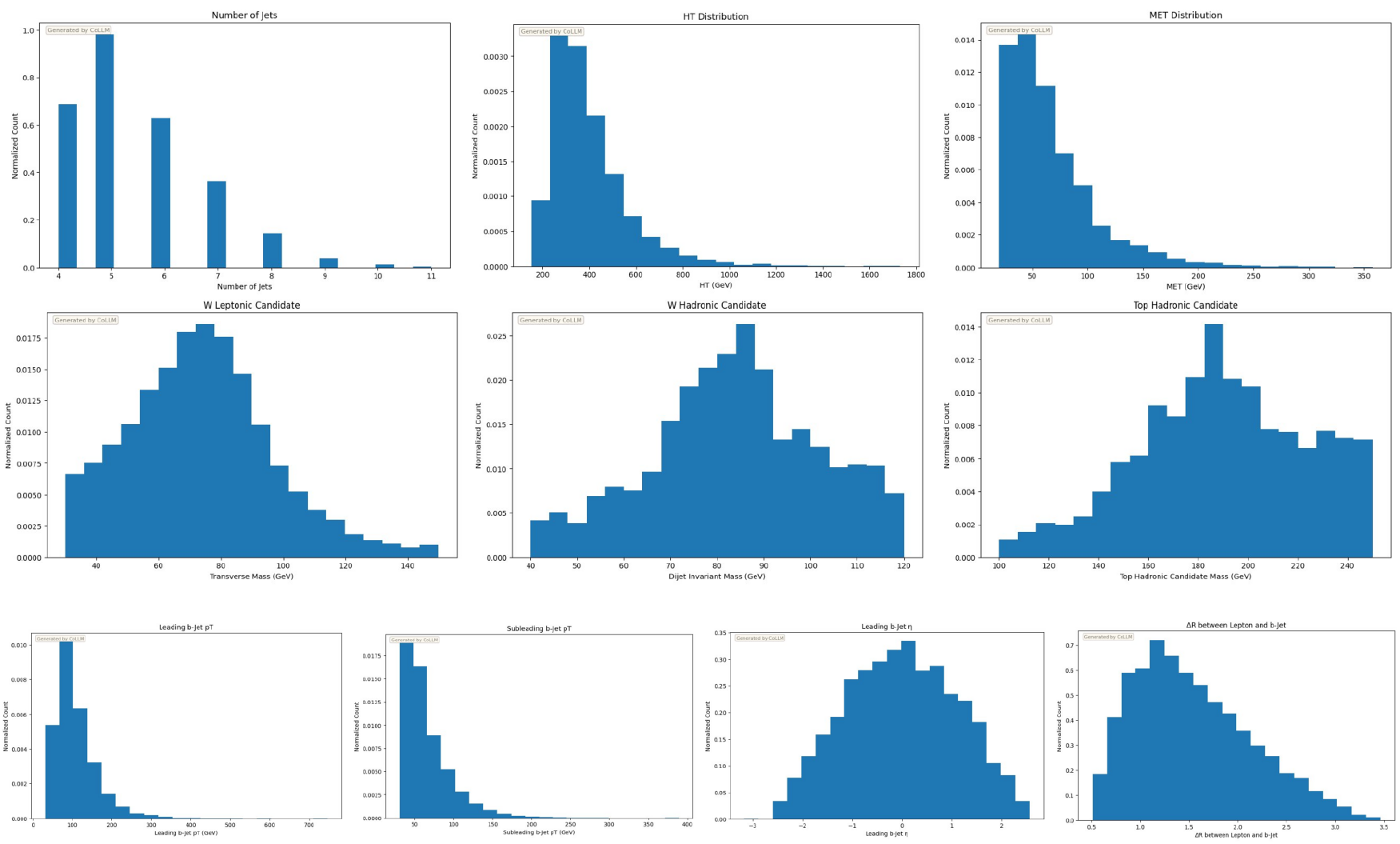}
    \caption{Validation histograms produced by \coll for the semi-leptonic $t\bar{t}$ analysis generated by \coll. The distributions show (top row) jet multiplicity, $H_{\mathrm{T}}$ and missing transverse energy; (middle row) transverse mass, $W$ hadronic candidate mass, and top hadronic candidate mass; (bottom row) leading and subleading $b$-jet $p_{\mathrm{T}}$, leading $b$-jet $\eta$, and $\Delta R$ between the lepton and closest $b$-jet. All histograms are normalized to unit area.}
    \label{fig:ttbar_plots}
\end{figure}

The generated  analysis code produces the requested  histograms by the user for physics validation, as shown in Fig.~\ref{fig:ttbar_plots}. The dijet invariant mass distribution peaks near the $W$ boson mass, indicating successful reconstruction of the hadronically decaying $W$. The invariant mass of the hadronic top candidate similarly peaks near the top quark mass. In addition, the transverse mass distribution of the lepton-MET system exhibits the characteristic Jacobian edge expected from $W \to \ell\nu$ decays.

To proceed to the deep learning stage, the user is asked to run the generated code  on the signal and background samples. \coll is tuned to produce a selection code in the output directory with a unique name as \texttt{generated\_lhco\_analysis.py}\footnote{The name of the generated code is fixed by \coll and always has the same name despite the selection analysis.}. This code accepts the input signal and background in LHCO format and can be run as
\begin{lstlisting}[
    basicstyle=\ttfamily\small,
    frame=single,
    backgroundcolor=\color{gray!10},
    language=bash
]
python generated_lhco_analysis.py full/path/to/.lhco 
\end{lstlisting}
The resulting CSV files are then supplied, via full file paths, to the machine learning configuration. The training stage uses these CSV inputs to construct a signal-to-background classifier. Users may specify the network architecture through either the graphical interface or a configuration file, as discussed in the previous subsections.

\section{Automated analysis pipeline}  
\label{sec:4}

This section describes the internal design of the LLM and deep learning classifiers, with particular emphasis on how explicit domain knowledge, deterministic code generation and execution feedback are combined to ensure robustness of the collider analysis. We first introduce the code generation engine, then discuss the automatic error detection and correction mechanism, and finally describe the automated deep learning classifiers.
\subsection{LangChain orchestrator and generation pipeline}
\label{subsubsec:generation_pipeline}

The LLM code generation engine constitutes the central  component of \coll. Its objective is not limited to producing isolated code fragments, but rather to generate complete and executable collider analysis workflows that closely resemble those written by experienced phenomenologists. These workflows include reading event level data (LHCO files), reconstructing physics objects, applying event selections, computing kinematic observables, producing histograms and exporting results in a structured format for deep learning analysis.

At the core of this process lies a LangChain orchestration layer, which is responsible for coordinating the interaction between multiple LLMs, enforcing physics constraints and managing the sequential stages of code generation and validation. LangChain helps break code generation into small steps, which allows \coll to treat code generation as multi-step task rather than a single step prompt response.
Concretely, the LangChain orchestration layer is responsible for
\begin{itemize}
    \item \textbf{Connecting to different LLMs}, providing a common interface that hides differences in tokenization, input limits and access methods;
    \item \textbf{Building the prompts}, combining system instructions, physics rules and user inputs into a well defined prompt;
    \item \textbf{Managing multistep generation}, breaking the task into planning, code generation and validation steps that are executed in sequence;
    
    \item \textbf{Applying constraints}, enforcing basic physics and formatting rules and using execution errors to guide automatic corrections;
    \item \textbf{Ensuring reproducibility.}
Reproducibility is enforced by using fully deterministic decoding during LLM inference. At each generation step $t$, the language model defines a probability distribution over the vocabulary,
\begin{equation}
p(w_t \mid w_{<t}) = \mathrm{softmax}\!\left(\frac{z_t}{T}\right),
\end{equation}
where $z_t$ denotes the model logits and $T$ is the temperature parameter. In \coll, the temperature is fixed to $T=0$, which suppresses stochasticity and collapses the distribution onto the most probable token.

Token selection is then performed via greedy decoding,
\begin{equation}
w_t = \arg\max_{w} \; p(w \mid w_{<t}),
\end{equation}
rather than sampling from the distribution. As a results, sampling ($\texttt{top\_p}=1$), top-$k$ filtering ($\texttt{top\_k}=0$), and probabilistic sampling ($\texttt{do\_sample}=\texttt{False}$) are disabled.

Under this configuration, the generation process becomes a deterministic function of the input prompt and model parameters. As a result, identical user input should produce similar analysis code across runs, which is essential for validation.

\end{itemize}
Unlike generic LLM, \coll formulates code generation as a constrained problem, guided by explicit physics rules and execution requirements. These constraints are enforced at the orchestration level, rather than being left implicit to the language model.
The orchestration pipeline employs two distinct large language models with complementary responsibilities. The first model is responsible for high level code generation, translating the structured analysis specification into an ordered sequence of computational steps. The second model focuses on low level code realization, resolving syntactic and implementation errors that may arise during code execution.

To ensure reproducibility and stability, the first model is configured to operate in a deterministic regime. In contrast, the second model is operated in a more exploratory mode to facilitate robust error correction. Specifically, error fixing is performed using temperature decoding with $T = 0.9$ and \texttt{do\_sample=True}, enabling the model to explore diverse strategies for resolving errors in the generated code.
This separation of responsibilities, combined with differentiated decoding configurations, allows \coll to improve the robustness and reproducibility of the end-to-end analysis pipeline.

\subsubsection{Physics aware system prompt}
\label{subsubsec:system_prompt}

In collider physics analyses, even minor deviations from established conventions, such as sign choices in four momentum summation, particle identification codes or reference mass values, can invalidate an entire analysis chain. To mitigate this risk, \coll employs a comprehensive, physics aware system prompt that encodes expert knowledge directly into the generation context.
This prompt acts as a structured prior on the generation process, constraining the hypothesis space explored by the language model and biasing it toward physically consistent code. The currently deployed system prompt is provided in \texttt{source/configs/system\_prompt.txt} and encodes the components described below.
It begins by defining the role and primary objectives of the LLM agent
\begin{lstlisting}[style=prompt]
You are a particle physics analysis assistant specialized in analyzing 
LHCO (.lhco) files produced by fast detector simulations (e.g., Delphes).

HARD REQUIREMENTS (must always be satisfied):
1. Always assume the input data is an LHCO file.
2. Always generate runnable Python 3 code when analysis is requested.
3. Always include the LHCO parser provided below.
4. Prioritize physics correctness over style or verbosity.
5. Use only: standard library, math, numpy (and matplotlib only if 
   explicitly needed).

PRIMARY TASK:
Given user input describing a physics goal, generate a correct LHCO-based 
analysis:
- Parse events and reconstructed objects
- Apply selection cuts
- Reconstruct physics objects (Z, W, H, top, etc.)
- Compute kinematic observables (pT, invariant mass, delta_R, MT)
- Output clear numerical results or histograms
\end{lstlisting}
These requirements ensure that the generated code remains self-contained and compatible with standard analysis.
For parsing the LHCO file, the system prompt provides a complete specification of the LHCO input file as
\begin{lstlisting}[style=prompt]
LHCO FILE FORMAT SPECIFICATION:

OBJECT LINE FORMAT:
  index  type  eta  phi  pt  jmass  ntrk  btag  had/em

COLUMN DEFINITIONS:
  index  : Object index within the event (0 marks new event header)
  type   : Particle type code (see below)
  eta    : Pseudorapidity
  phi    : Azimuthal angle (radians)
  pt     : Transverse momentum (GeV)
  jmass  : Jet mass (GeV) -- use only for jets
  ntrk   : Track count; sign encodes lepton charge
  btag   : b-tag flag (1.0 = b-tagged jet, 0.0 = not b-tagged)
  had/em : Hadronic-to-electromagnetic energy ratio

PARTICLE TYPE CODES:
  0 = Photon
  1 = Electron
  2 = Muon
  3 = Tau
  4 = Jet
  6 = MET (eta = 0, phi = MET direction, pt = MET magnitude)
\end{lstlisting}
This explicit encoding prevents common errors, such as misidentification of MET type codes or incorrect determination of lepton charge.
A complete and validated parser is embedded in the system prompt to ensure consistent event handling across all generated analyses as 
\begin{lstlisting}[style=pythonstyle]
def read_lhco(filename):
    """Parse LHCO file and return list of events."""
    events = []
    current = None
    
    with open(filename) as f:
        for line in f:
            line = line.strip()
            if not line or line.startswith('#'):
                continue
            
            parts = line.split()
            
            # New event header (index = 0)
            if parts[0] == '0':
                if current is not None:
                    events.append(current)
                current = {'id': int(parts[1]), 'objects': []}
            
            # Object line
            elif current is not None:
                obj = {
                    'type':   int(parts[1]),
                    'eta':    float(parts[2]),
                    'phi':    float(parts[3]),
                    'pt':     float(parts[4]),
                    'jmass':  float(parts[5]),
                    'ntrk':   float(parts[6]),
                    'btag':   float(parts[7]) if len(parts) > 7 else 0.0,
                    'had_em': float(parts[8]) if len(parts) > 8 else 0.0
                }
                # Derive charge from ntrk sign
                obj['charge'] = 1 if obj['ntrk'] > 0 else \
                               (-1 if obj['ntrk'] < 0 else 0)
                current['objects'].append(obj)
    
    if current is not None:
        events.append(current)
    
    return events
\end{lstlisting}
The parser handles event boundaries, skips comments and empty lines and automatically derives particles properties according to the LHCO convention.

To reconstruct the kinematic variables, the system prompt defines the rules for calculating these quantities. In the current implementation, 43 kinematic variables are defined. For example, the invariant mass formula is explicitly specified, along with warnings to prevent a common mistake often seen in code generated by LLM as

\begin{lstlisting}[style=prompt]
INVARIANT MASS OF N PARTICLES:
**CRITICAL: Sum the 4-momentum components (NOT differences!), then compute:**

    E_total  = E_1 + E_2 + ... + E_i
    px_total = px_1 + px_2 + ... + px_i
    py_total = py_1 + py_2 + ... + py_i
    pz_total = pz_1 + pz_2 + ... + pz_i

    M^2 = E_total^2 - px_total^2 - py_total^2 - pz_total^2
\end{lstlisting}
Example definitions of transverse mass and angular separation are
\begin{lstlisting}[style=prompt]
TRANSVERSE MASS (MT):

    MT = sqrt(2 * pT_visible * MET * (1 - cos(delta_phi)))

ANGULAR SEPARATION (delta_R):

    delta_eta = eta_1 - eta_2
    delta_phi = phi_1 - phi_2  (normalized to [-pi, pi])
    delta_R = sqrt(delta_eta^2 + delta_phi^2)
\end{lstlisting}

Moreover, the system prompt includes a set of reference helper functions for particle selection and sorting:
\begin{lstlisting}[style=pythonstyle]
def get_leptons(event):
    """Return all electrons and muons."""
    return [o for o in event['objects'] if o['type'] in (1, 2)]

def get_bjets(event):
    """Return all b-tagged jets."""
    return [o for o in event['objects'] if o['type'] == 4 and o['btag'] == 1.0]

def get_met(event):
    """Return MET object or None."""
    for o in event['objects']:
        if o['type'] == 6:
            return o
    return None

def sort_by_pt(particles):
    """Return particles sorted by pT (descending)."""
    return sorted(particles, key=lambda p: p['pt'], reverse=True)
\end{lstlisting}
These implementations establish consistent naming conventions and interface patterns that guide subsequent code generation.
In addition to physics aware content, the prompt encodes best practices for event based analyses as the following:
\begin{lstlisting}[style=prompt]
BEST PRACTICES:
- Keep code minimal, explicit, and readable
- Validate particle counts before pairing or selection
- Handle missing objects and malformed inputs gracefully
- Re-extract particle collections inside each event loop
- MET type code is 6, NOT 5
- Cutflow outputs must match applied cuts
\end{lstlisting}
Finally, output and clarification procedures are also specified.
\begin{lstlisting}[style=prompt]
OUTPUT RULES:
- Output code only unless explanation is requested
- No placeholder logic
- State assumptions in comments when needed

CLARIFICATION POLICY:
If the request is ambiguous:
- Adopt a standard analysis strategy
- State assumptions clearly
- Proceed without interrupting the workflow
\end{lstlisting}
By embedding format specifications, kinematic formulas, reference implementations and physics constraints directly into the generation context, \coll minimizes ambiguity and reduces reliance on potentially inconsistent information acquired during code generation. We find that the current prompt structure is well suited for the majority of standard collider analyses.
However, for more complex analysis, users may extend or customize the system prompt by incorporating additional  functions, constants or conventions. 

\subsubsection{Automatic error correction}
\label{subsec:pyfixer}

Despite careful prompt engineering and the use of large scale language models, automatically generated analysis code may still fail at execution time. Such failures typically arise from minor syntactic errors, missing imports or incorrect assumptions about data structures.  To mitigate this limitation, \coll incorporates an automatic error correction module, \texttt{PyFixer}, which enables self corrected code execution. Instead of relying on user driven debugging, \texttt{PyFixer} implements an iterative repair mechanism that analyzes execution failures and guides the language model toward corrective modifications. This design is motivated by the fact that LLMs are generally more effective at refining and correcting existing codes.

Error correction is performed through an iterative refinement loop. At each iteration, the current code version, annotated with line numbers, is passed to the LLM together with the repair prompt. Rather than regenerating the entire script, the model is instructed to modify only the faulty sections. This localized editing strategy preserves validated logic in unaffected regions and accelerates convergence toward a correct solution. The loop terminates when executable code is obtained or when a predefined maximum number of iterations is reached. In practice, the majority of execution failures are resolved within one or two iterations, demonstrating the effectiveness of the refinement loop. The behavior of the repair process is governed by a dedicated system prompt that instructs the LLM to perform  comprehensive debugging as follows
\begin{lstlisting}[style=prompt]
You are an expert Python debugger. Your task is to fix buggy Python code.
Process:
1. Read the error message carefully - note the error TYPE and LINE NUMBER 
2. Understand what the code is trying to do 
3. Identify the root cause (not just the symptom)
4. Fix ALL issues, not just the reported one
5. Return COMPLETE working code
Common fixes:
- SyntaxError: missing colons, parentheses, commas, quotes
- NameError: typos in variable names, undefined variables, missing imports
- TypeError: wrong types, missing/extra arguments, None operations
- IndexError: check list bounds, empty lists
- KeyError: check dict keys exist, use .get()
- AttributeError: wrong method names, None objects
- IndentationError: fix spacing consistency
- ImportError: correct module names, install missing packages
Rules:
- Return the COMPLETE fixed code, not just the changed parts
- Keep all original functionality
- Add missing imports if needed
- Use ```python``` code block
- NO explanations outside the code block 
\end{lstlisting}
%

\subsubsection{LLM selection and Trade-offs}
\label{subsubsec:model_selection}

The quality of the generated analysis code depends critically on the underlying large language model. \coll therefore supports a wide range of  LLMs that have demonstrated strong performance on structured programming and scientific computing tasks. The supported models span a broad range of architectures and parameter scales, including lightweight models optimized for rapid iteration, medium scale models offering a favorable balance between performance and efficiency and large scale models designed for complex code generation. In particular, \coll supports recent models from the Qwen, Llama and DeepSeek families. Table~\ref{tab:llm_comparison} summarizes the main characteristics of the language models currently supported by \coll.

Medium size models, such as \texttt{Qwen2.5-Coder-32B-Instruct}, offer a good balance between output quality and computing cost. They are well suited for physics analyses on mid-range GPUs. Larger models, including \texttt{Llama-3.3-70B-Instruct}, provide better long range consistency and follow domain specific rules more reliably, but they require more memory and longer inference times. Smaller models are still useful for simple analyses or when computing resources are limited.

\begin{table}[!h]
\centering
\caption{Comparison of representative large language models supported by \coll. The approximate GPU memory requirements for inference without aggressive quantization are shown. The Llama-3.3-70B model is recommended; however, it requires an API key.}
\label{tab:llm_comparison}
\resizebox{\textwidth}{!}{%
\begin{tabular}{lccclc}
\hline
Model & Params & Category & Focus & Key features & Memory [GB] \\
\hline
Qwen2.5-Coder-3B \cite{qwen25}  & 3B  & Coder     & Code gen. & High speed, low footprint      & 6--8   \\
Llama-3.2-3B \cite{llama3}      & 3B  & General   & General   & Lightweight, stable            & 6--8   \\
Qwen2.5-Coder-7B \cite{qwen25}  & 7B  & Coder     & Code gen. & Balanced performance           & 14--16 \\
Llama-3.1-8B \cite{llama3}      & 8B  & General   & General   & Robust, versatile              & 16--18 \\
Qwen3-Coder-30B \cite{qwen3}   & 30B & Coder     & Code gen. & Scalable performance           & 55--65 \\
Qwen2.5-Coder-32B \cite{qwen25} & 32B & Coder     & Code gen. & High precision                 & 60--70 \\
DeepSeek-R1-32B \cite{deepseekr1}   & 32B & Reasoning & Analysis  & Multi-step inference           & 60--70 \\
QwQ-32B \cite{qwq}           & 32B & Reasoning & Planning  & Structured reasoning           & 60--70 \\
\textbf{Llama-3.3-70B} \cite{llama3}     & 70B & General   & General   & Long range coherence           & 130--150 \\
Qwen2.5-72B\cite{qwen25}       & 72B & General   & Analysis  & Strong analytical capability   & 135--155 \\
\hline
\end{tabular}}
\end{table}

By using a unified model interface  \coll allows users to switch easily between different language models without changing their usual workflow. This flexible design helps users choose the best balance between code quality, execution speed and available hardware. In addition, if users have an API key, \coll can connect to cloud, making it possible to use HuggingFace machines for inference.

\subsection{Automated deep learning analysis}
\label{subsec:dl_analysis}

Following the  selection code generation stage, in which cuts are applied and discriminating kinematic variables are extracted, the analysis pipeline proceeds to the deep learning classification phase. At this stage, neural network models are employed to construct optimized decision boundaries in the high dimensional feature space, thereby enhancing the separation between signal and background events. \coll provides three complementary network architectures, MLP, GNN and Transformers, each suited for different data structure of collider events.

The framework automates the construction, training and evaluation of these models, enabling users to deploy advanced deep learning classifiers without requiring extensive expertise in deep learning methodologies. The automated deep learning pipeline in \coll takes as input the CSV files produced from running the LLM generated code on full set of signal and background events. Users configure the network architecture through either graphical or terminal interfaces, specifying structural hyperparameters such as layer depth and width, regularization schemes and optimization algorithms.
Network training is carried out with configurable callbacks for early stopping, learning rate scheduling and checkpoint management. Upon completion, the pipeline produces trained model parameters, quantitative performance metrics evaluated with multiple figures of merit and diagnostic visualizations, including loss histories and classifier score distributions for signal and background events.

The selection among MLP, GNN and Transformer architectures depends on the analysis requirements and available computational resources. MLPs offer the fastest training, making them ideal for analyses where extracted high level features already capture relevant physics. GNNs excel when relational structure among particles carries discriminating information not easily encoded in kinematic variables. Transformers provide maximum flexibility and often achieve the highest performance, at the cost of increased computational demands and sensitivity to hyperparameter tuning.
\subsubsection{Multi-Layer Perceptron}
\label{subsubsec:mlp}

Despite the emergence of more sophisticated architectures, MLP remains the standard baseline for tabular kinematic data, offering robust performance across a wide range of physics analyses while requiring minimal architectural tuning.
An MLP comprises an input layer that receives the reconstructed kinematic features, one or more hidden layers that perform successive nonlinear transformations and an output layer that produces classification scores. The number of neurons in the input layer is automatically adjusted according to the size of the inputs, while the output layer always has two neurons. \coll supports three types of hidden layers, \texttt{Dense}, \texttt{Dropout} and \texttt{Batch Normalization} layers. Only dense layer receives input parameters and  computes the output as 
\begin{equation}
\mathbf{h}^{(l)} = \sigma\!\left( \mathbf{W}^{(l)} \mathbf{h}^{(l-1)} + \mathbf{b}^{(l)} \right),
\end{equation}
where $\mathbf{h}^{(l)}$ denotes the activation vector at layer $l$, $\mathbf{W}^{(l)}$ and $\mathbf{b}^{(l)}$ are the trainable weight matrix and bias vector, respectively, and $\sigma(\cdot)$ represents a non-linear activation function. The input layer receives the standardized feature vector $\mathbf{x} \in \mathbb{R}^d$, where $d$ denotes the number of kinematic variables constructed by the generated LLM selection code.

The choice of activation function plays a central role in determining the expressive power and optimization behaviour of the network. \coll supports a broad range of activation functions, with the Rectified Linear Unit (ReLU)  adopted as the default for hidden layers. ReLU, defined as $\sigma(z)=\max(0,z)$, promotes sparse activations and alleviates the vanishing gradient problem commonly encountered in deep networks with saturating nonlinearities. In scenarios where inactive ReLU units degrade performance, Leaky ReLU and its parametric variant PReLU \cite{he2015delving} preserve small gradients for negative inputs. The Exponential Linear Unit (ELU) \cite{clevert2015fast} and Scaled ELU (SELU) \cite{klambauer2017self} provide smoother behaviour around zero, with SELU designed to induce self-normalizing properties in deep architectures. The Gaussian Error Linear Unit (GELU) \cite{hendrycks2016gaussian}, originally introduced in the context of transformer models, has also proven effective in MLP classifiers. 

The depth and width of the network jointly determine its classification performance. Shallow architectures with wide hidden layers efficiently capture global feature correlations but may be limited in modelling hierarchical structures. Conversely, deeper networks with narrower layers can learn compositional representations at the cost of increased optimization complexity. Empirical studies in collider analyses indicate that architectures comprising two to four hidden layers with 64 to 256 neurons per layer are sufficient for most classification tasks involving $\mathcal{O}(10)$ input features. The \coll interface facilitates systematic architecture exploration, allowing users to incrementally construct networks and configure layer specific parameters. These configurations can be adopted from the input YAML file as shown in subsection \ref{subsubsec:tui}.

\subsubsection{Graph Neural Networks}
\label{subsubsec:gnn}

In contrast to MLP classifiers, which operate on fixed dimensions feature vectors, GNNs process sets of particles and their pairwise relations directly. This enables the automatic discovery of physically relevant interaction patterns without the need for manually constructed composite observables.
A collider event admits natural graph representations at multiple levels of granularity. Each reconstructed particle such as electrons, muons, jets, photons and missing transverse energy, is represented as a node, while edges encode potential interactions between object pairs \cite{Qu:2019gqs,Esmail:2024jdg}. Node features describe the kinematic properties of each particle. A minimal feature set consists of the transverse momentum $p_T$, pseudorapidity $\eta$, azimuthal angle $\phi$, and invariant mass $m$, supplemented by derived quantities such as the energy $E$ and momentum components $(p_x,p_y,p_z)$. In \coll, node features are standardized by subtracting the mean and dividing by the standard deviation computed over the training events.

The construction of edges determines which particle pairs may exchange information during message passing. Fully connected graphs, in which all nodes are mutually linked, maximize information flow but introduce $\mathcal{O}(n^2)$ edges for $n$ particles. In the current implementation, \coll adopts fully connected graphs with constant edge features equal to unity, providing a simple and general baseline for relational modeling. We keep the structure of more advanced graph to future versions of \coll.

Message passing constitutes the core computational mechanism of GNNs. Each layer updates node representations by aggregating information from neighboring nodes, enabling the progressive integration of local and global event features. The generic message passing scheme can be expressed as
\begin{align}
\mathbf{m}_{ij}^{(l)} &= \phi_e\!\left( \mathbf{h}_i^{(l)}, \mathbf{h}_j^{(l)}, \mathbf{e}_{ij} \right), \\
\mathbf{h}_i^{(l+1)} &= \phi_h\!\left( \mathbf{h}_i^{(l)}, \bigoplus_{j \in \mathcal{N}(i)} \mathbf{m}_{ij}^{(l)} \right),
\end{align}
where $\mathbf{h}_i^{(l)}$ denotes the hidden representation of node $i$ at layer $l$, $\mathbf{e}_{ij}$ encodes the edge features between nodes $i$ and $j$, $\phi_e$ and $\phi_h$ are learnable functions (typically a fully connected layer), $\mathcal{N}(i)$ denotes the neighborhood of node $i$ and $\bigoplus$ represents a permutation invariant aggregation operator, such as summation, averaging or max pooling.

Several GNN architectures implement this framework using different design choices. GCNs employ normalized adjacency matrices and simplified aggregation rules. GATs \cite{velickovic2018graph} assign dynamic, learned weights to edges through attention mechanisms, allowing the model to focus on physically relevant interactions. EdgeConv \cite{wang2018dynamic}, constructs edge features from differences between node features and performs convolutions in this edge space.

After $L$ message passing layers, the node representations contain both local kinematic information and information from their surrounding neighborhoods. A global readout operation then combines all node information into a single vector that represents the whole event,
\begin{equation}
\mathbf{h}_{\text{graph}} = \bigoplus_{i \in \mathcal{V}} \mathbf{h}_i^{(L)},
\end{equation}
where $\mathcal{V}$ denotes the set of nodes. More elaborate readout schemes, including attention-based pooling and hierarchical clustering, may be employed to capture multi-scale event structure. The resulting graph embedding is passed to a classification head.

A key feature of GNNs is their invariance under permutations of node ordering. Since physical observables must be independent of arbitrary indexing conventions, this property is essential for collider analyses. By construction, GNN aggregation operators enforce permutation invariance, obviating the need for explicit data augmentation strategies.

In \coll, GNN architectures are implemented through integration with PyTorch, which provides optimized message passing layers and utilities for graph structured data. Users specify the number of message passing layers, hidden dimensions, activation functions, normalization schemes, pooling strategies and output heads through a structured configuration file.
An example configuration for a GCN is shown below
\begin{lstlisting}[language=YAML]
#===========================
#Model Architecture - GCN
#===========================
model:
  type: GCN
  layers:
    - out_channels: 64
      activation: relu
      batchnorm: true
      dropout: 0.1

    - out_channels: 64
      activation: relu
      improved: false
      batchnorm: true
      dropout: 0.1

    - out_channels: 64
      activation: relu
      improved: false
      batchnorm: false
      dropout: 0.0

  pooling: global_mean         # global pooling method
  output_units: 2              # binary classification
  output_activation: null      # null for logits
\end{lstlisting}

In this configuration, each entry in the \texttt{layers} field defines a message passing layer and specifies its output dimensionality, activation function, normalization and regularization settings. The \texttt{pooling} field controls the global readout operation, while \texttt{output\_units} and \texttt{output\_activation} define the classification head. 
\paragraph{Dynamic Edge Convolution Networks:}
\label{subsubsec:edgeconv}

While GCNs operate on graphs with fixed connectivity, as discussed in the previous subsection, EdgConv introduces a more flexible framework in which graph edges are updated dynamically during training. In this approach, the EdgeConv operation constitutes the fundamental building block and treats edges, rather than nodes alone.
For an edge connecting nodes $i$ and $j$, the EdgeConv operation is defined as
\begin{equation}
\mathbf{h}_{ij}^{(l)} = \Phi^{(l)}\!\left( \mathbf{v}_i^{(l)}, \mathbf{v}_j^{(l)} - \mathbf{v}_i^{(l)} \right),
\end{equation}
where $\mathbf{v}_i^{(l)}$ denotes the feature vector of node $i$ at layer $l$, and $\Phi^{(l)}$ is a fully connected layer applied to the concatenation of the central node features and relative feature differences. This construction enables the network to learn local relational patterns. The updated representation of node $i$ is obtained by aggregating messages from its neighborhood,
\begin{equation}
\mathbf{v}_i^{(l+1)} =
\rho \left( \left\{ \mathbf{h}_{ij}^{(l)} \mid j \in \mathcal{N}_k(i) \right\} \right),
\end{equation}
where $\mathcal{N}_k(i)$ denotes the set of $k$ nearest neighbors of node $i$ in the learned feature space, and $\rho$ is a permutation invariant aggregation function, such as maximum or mean pooling.

In contrast to GCNs, where the adjacency structure is fixed throughout training, the neighborhood $\mathcal{N}_k(i)$ in EdgConv is recomputed at every layer using the updated node representations. As a result, the effective graph topology evolves dynamically, allowing the model to progressively focus on physically relevant object correlations. 
After several EdgeConv layers, the node representations encode both local geometric information and higher order correlations. A global pooling operation then combines these representations into a single vector, which is passed to a classification head for final signal to background discrimination.

Users specify the network structure through a YAML configuration file, which controls the number of layers, neighborhood size, aggregation method and regularization options. An example configuration is shown below

\begin{lstlisting}[language=YAML]
#===============================
#Model Architecture - EdgeConv
#===============================
model:
  type: EdgeConv
  layers:
    - out_channels: 64
      activation: relu
      k: 7                    # Number of nearest neighbors
      aggr: max               # Aggregation: max, mean, add
      batchnorm: true
      dropout: 0.1
    - out_channels: 64
      activation: relu
      k: 7
      aggr: max
      batchnorm: true
      dropout: 0.1
    - out_channels: 64
      activation: relu
      k: 7
      aggr: max
      batchnorm: false
      dropout: 0.0

  pooling: global_max         # Global pooling method
  output_units: 2             # Binary classification
  output_activation: null     # Logits
\end{lstlisting}

In this configuration, each layer defines the dimensionality of the node embeddings through \texttt{out\_channels}, the neighborhood size through \texttt{k} and the aggregation strategy through \texttt{aggr}. Batch normalization and dropout can be enabled in each layer to improve training stability and reduce overfitting. The \texttt{pooling} option specifies how node features are combined at the graph level.

The modular structure of the configuration file allows users to extend the architecture to arbitrary depths and to incorporate additional layer types or customized message passing blocks. 
\paragraph{Graph Attention Networks:}

GATs extend the GCN by incorporating attention mechanisms that adaptively weight the contributions of neighboring nodes during feature aggregation~\cite{Esmail:2023axd,Fuks:2025qgh}. In contrast to standard GCNs, which assign fixed or uniformly normalized weights based on graph topology, GATs learn  importance scores that enable the model to focus on the most relevant particle  correlations for the classification task. By allowing the network to emphasize informative interactions and suppress less relevant ones, GATs provide a more expressive representation of event structure.

The attention mechanism in GATs assigns a learnable coefficient to each pair of connected nodes. For a node $i$ and one of its neighbors $j$, the unnormalized attention score is defined as
\begin{equation}
e_{ij} =
\text{LeakyReLU}
\left(
\mathbf{a}^{\top}
\left[
\mathbf{W}\mathbf{h}_i
\oplus
\mathbf{W}\mathbf{h}_j
\right]
\right),
\end{equation}
where $\mathbf{h}_i$ and $\mathbf{h}_j$ denote the input feature vectors of nodes $i$ and $j$, respectively, $\mathbf{W} \in \mathbb{R}^{F' \times F}$ is a shared linear transformation, $\oplus$ denotes concatenation and $\mathbf{a}$ is a learnable attention vector. The LeakyReLU activation improves numerical stability and gradient flow.
The attention scores are normalized over the neighborhood of node $i$ using the softmax function,
\begin{equation}
\alpha_{ij}
=
\frac{\exp(e_{ij})}
{\sum_{k \in \mathcal{N}(i)} \exp(e_{ik})},
\end{equation}
where $\mathcal{N}(i)$ denotes the set of neighbors of node $i$. This normalization ensures that the attention weights form a convex combination and are invariant under permutations of node ordering.

The updated representation of node $i$ is then obtained by a weighted aggregation of its neighbors,
\begin{equation}
\mathbf{h}_i^{(l+1)}
=
\sigma
\left(
\sum_{j \in \mathcal{N}(i)}
\alpha_{ij}\,
\mathbf{W}\mathbf{h}_j^{(l)}
\right),
\end{equation}
where $\sigma(\cdot)$ is a nonlinear activation function. This operation allows each node to adaptively combine information from its neighborhood based on learned relevance scores.

To enhance model capacity and stabilize training, GATs commonly employ multi-head attention, in which several independent attention mechanisms operate in parallel. Their outputs are either concatenated or averaged, providing a richer representation of local interactions.

After several attention layers, the node embeddings encode both kinematic information and learned relational features. A   pooling operation then aggregates these representations into a single vector, which is passed to a classification head for final discrimination.
Users specify the network structure through a YAML configuration file, which controls the number of attention heads, feature dimensions, normalization, and regularization options. An example configuration is shown below:

\begin{lstlisting}[language=YAML]
#===========================
#Model Architecture - GAT
#===========================
model:
  type: GAT
  layers:
    - out_channels: 32
      activation: relu
      heads: 4                 # Number of attention heads
      concat: true             # Concatenate head outputs
      dropout: 0.1             # Attention dropout
      batchnorm: true
    - out_channels: 32
      activation: relu
      heads: 4
      concat: true
      dropout: 0.1
      batchnorm: true
    - out_channels: 64
      activation: relu
      heads: 1                 # Final layer: single head
      concat: false            # No concatenation
      dropout: 0.0
      batchnorm: false

  pooling: global_mean         # Graph-level pooling
  output_units: 2              # Binary classification
  output_activation: null      # Logits
\end{lstlisting}

In this configuration, the \texttt{heads} parameter controls the number of parallel attention mechanisms in each layer, while \texttt{concat} specifies whether their outputs are concatenated or averaged. The \texttt{out\_channels} parameter determines the dimensionality of each head. Dropout and batch normalization can be enabled to improve generalization and training stability.

\subsubsection{Transformer Networks}
\label{subsubsec:transformer}

Transformer architectures, originally developed for natural language processing \cite{vaswani2017attention}, have recently demonstrated strong performance in collider physics applications \cite{Qu:2022mxj,Hammad:2023sbd,Hammad:2024qme,Esmail:2025kii,Mikuni:2024qsr,Spinner:2024hjm,Brehmer:2024yqw,Wu:2024thh,He:2023cfc}. Their core self-attention mechanism enables the direct modeling of long range correlations among reconstructed particles, without relying on the locality constraints inherent to convolutional or message-passing approaches. In transformer  analyses, events are represented as unordered sets of particle embeddings \cite{Komiske:2018cqr,Qu:2019gqs}. Each reconstructed object is first mapped into a $d$-dimensional latent space through a learned linear projection,
\begin{equation}
z_i^{(0)} =
W_{\mathrm{embed}} x_i
+
\mathbf{b}_{\mathrm{embed}},
\end{equation}
where $x_i$ denotes the input feature vector of particle $i$. Angular and geometric information is instead incorporated directly through the input features. The self-attention operation computes pairwise interactions between all particles in an event. Given the embedding matrix $Z = [z_1,\ldots,z_n]$, the model constructs query, key, and value matrices.
\begin{equation}
Q = Z\ W_Q,
\qquad
K = Z\ W_K,
\qquad
V =Z\ W_V,
\end{equation}
where $W_Q$, $W_K$, and $W_V$ are learnable projection matrices. The attention output is then given by
\begin{equation}
\mathrm{Attention}(Q,K,V)
=
\mathrm{softmax}
\left(
\frac{Q \ K^{\top}}{\sqrt{d_k}}
\right)
V.
\end{equation}
The softmax normalization ensures that attention weights sum to unity, while the factor $\sqrt{d_k}$ stabilizes gradients for large embedding dimensions. Through this operation, each particle aggregates information from all other particles, weighted by learned relevance scores.

To increase representational capacity, transformers employ multi-head attention, in which several independent attention mechanisms operate in parallel. Their outputs are concatenated and linearly projected,
\begin{equation}
\mathrm{MultiHead}(Z)
=
\mathrm{Concat}(\mathrm{head}_1,\ldots,\mathrm{head}_h) W_O.
\end{equation}

A transformer layer combines multi-head attention with a  feed-forward network and residual connections as,
\begin{align}
&Z' = \mathrm{LayerNorm} \left( Z+ \mathrm{MultiHead}(Z)\right), \\
&Z^{(l+1)} = \mathrm{LayerNorm} \left(Z'+ \mathrm{FFN}(Z')\right),
\end{align}
where $\mathrm{FFN}$ is a two-layer MLP with GELU activation. Layer normalization and residual connections improve training stability and enable the construction of deep architectures.
After a stack of transformer layers, the particle embeddings are aggregated into a fixed-dimensional event representation. This can be achieved through mean pooling or attention-weighted pooling. The resulting  vector is processed by a classification head to produce signal and background scores.

Standard self-attention scales as $\mathcal{O}(n^2)$ with the number of particles $n$, which can limit applicability for very high multiplicity events. However, for typical collider applications with $\mathcal{O}(10$--$100)$ particles, this cost remains manageable on modern GPUs. 
Training transformers generally requires smaller learning rates than MLPs or GNNs, typically in the range $10^{-6}$--$10^{-4}$, combined with warmup schedules. Regularization through dropout and weight decay is essential to control overfitting given the large parameter counts.

In \coll, transformer architectures are configured using a YAML file that specifies embedding dimensions, attention structure, and regularization parameters. An example configuration is shown below

\begin{lstlisting}[language=YAML]
#==================================
#Model Architecture - Transformer
#==================================
model:
  embed_dim: 128                # Particle embedding dimension
  num_heads: 8                  # Number of attention heads
  num_layers: 4                 # Transformer layers
  ffn_dim: 256                  # Feed-forward network dimension
  dropout: 0.1                  # Dropout rate
  attention_dropout: 0.1        # Attention dropout
  pooling: mean                 # mean, max, attention, cls
  num_classes: 2                # Binary classification
  pre_norm: true                # Pre-layer normalization
\end{lstlisting}

Here, \texttt{embed\_dim} controls the latent feature dimension, while \texttt{num\_heads} and \texttt{num\_layers} determine the depth and parallelism of the attention mechanism. The \texttt{pooling} option selects the method used to construct the event representation.

\subsubsection{Data structure and hyperparameter tuning}
\label{subsubsec:data_hyper}

All deep learning models in \coll\ accept input data in comma separated values CSV format. However, the internal data representation depends on the selected architecture. For MLPs, each row in the input file corresponds to a single event, and each column represents a reconstructed kinematic feature. The number of neurons in the input layer is automatically adjusted according to the number of columns in the input CSV file. Users may specify the desired set of features during the LLM code generation stage as
\begin{lstlisting}[style=prompt]
[OUTPUT_STRUCTURE] 
- Save the following in a CSV file for MLP analysis:
  1- Transverse mass of lepton + MET
  2- Dijet invariant mass
  3- MET
  4- pT of the lepton
  5- Delta R between the two jets
  .....
  .....
\end{lstlisting}
This specification automatically configures the output of the generated selection code for subsequent MLP analysis.

In contrast, graph neural networks and transformer models operate on particle sets representations. In these cases, each row in the input file corresponds to a single reconstructed particle, and each column encodes the associated kinematic or identification features. Events are represented as collections of particles, forming either graphs or particle clouds. For both architectures, users must specify the number of particles per event. The framework then iterates over the CSV file and groups rows accordingly to construct individual events. The input format can be defined during the LLM code generation stage as 
\begin{lstlisting}[style=prompt]
[OUTPUT_STRUCTURE] 
- Save the following in a CSV file for GNN analysis:
  Save the following node features: [pt, eta, phi, charge]
  1- Electron
  2- Muon
  3- Leading jet
  4- Subleading jet
  .....
  .....
\end{lstlisting}

As fully connected graph models and particle transformer architectures share the same data structure. Consequently, the above output specification can be used for both GNN and transformer analyses.

%
The model setup depends on user choices, and users are encouraged to apply additional methods to stabilize training, especially when working with limited data. To reduce overfitting and improve generalization, regularization techniques can be used. Dropout layers can be added between hidden layers to randomly deactivate some neurons during training, helping the model learn more robust features rather than relying on individual neurons. Batch normalization is recommended to normalize layer inputs within each minibatch, making training faster and more stable. In addition, weight decay (also known as L2 regularization) helps limit large parameter values and encourages smoother and more reliable decision boundaries.
Model parameters are optimized by minimizing the binary cross-entropy loss,
\begin{equation}
\mathcal{L} = -\frac{1}{N} \sum_{i=1}^{N} \left[ y_i \log p_i + (1 - y_i) \log(1 - p_i) \right],
\end{equation}
where $y_i \in {0,1}$ denotes the true class label and $p_i$ is the predicted signal probability for event $i$.
Optimization is performed using variants of stochastic gradient descent. The available optimizers include \texttt{adam}, \texttt{adamw} and \texttt{sgd}. For Transformer-based architectures, AdamW is used as the default due to its decoupled treatment of weight decay, whereas Adam is used for MLP and GNN models. These optimizers provide adaptive learning rates and momentum terms, leading to improved convergence and numerical stability.

Learning rate schedulers are used to dynamically adjust the step size during training. The supported scheduler types are: \texttt{none}, corresponding to a constant learning rate; \texttt{step}, which reduces the learning rate by a factor \texttt{gamma} every \texttt{step\_size} epochs; \texttt{plateau}, which is triggered by saturation of the validation loss and controlled by the parameter \texttt{patience}; \texttt{cosine}, which follows a cosine annealing schedule; \texttt{cosine\_warmup}, which implements cosine annealing with an initial warmup phase; and \texttt{onecycle}, which combines warmup and decay phases with a specified \texttt{max\_lr}.

Early stopping is employed to further suppress overfitting by monitoring either the validation loss, denoted by \texttt{val\_loss}, or a selected evaluation metric, denoted by \texttt{eval\_metric} and terminating training when no further improvement is observed. The maximum number of consecutive epochs without improvement is controlled via \texttt{early\_stopping\_patience}. Supported evaluation metrics include \texttt{accuracy}, \texttt{auc}, \texttt{f1}, \texttt{recall}, and \texttt{precision}.
Training precision can be configured as \texttt{float32}, corresponding to full precision; \texttt{float16}, corresponding to half precision; or \texttt{mixed}, corresponding to automatic mixed precision, allowing a trade-off between numerical accuracy and computational efficiency. In addition, gradient clipping, controlled by the parameter \texttt{gradient\_clip}, is available to constrain gradient magnitudes and stabilize training, which is particularly beneficial for Transformer models.
An example configuration of the data preprocessing and training setup is presented below.
\noindent 
\

\begin{lstlisting}[language=YAML]
#======================
# Data Configuration
#======================
data:
  signal_path: "data/signal_GNN.csv"
  background_path: "data/bkg_gnn.csv"
  train_size: 1000        # samples per class for training
  test_size: 1000         # samples per class for testing
  val_ratio: 0.15         # fraction of training data for validation
  normalize: true         # normalize features with StandardScaler

#======================
# Training Configuration
#======================
train:
  epochs: 50
  batch_size: 64
  learning_rate: 0.0001
  weight_decay: 0.01      # L2 regularization
  optimizer: adamw        # Options: adam, adamw, sgd
  device: auto            # Options: auto, cpu, cuda, cuda:0, cuda:1

  # Early Stopping
  early_stopping: true
  early_stopping_patience: 10
  early_stopping_metric: val_loss  # Options: val_loss, eval_metric

  # Training Precision
  precision: float32      # Options: float32, float16, mixed

  # Gradient Clipping
  gradient_clip: 1.0

  # Learning Rate Scheduler
  scheduler:
    type: cosine          # Options: none, step, plateau, cosine, 
                          #          cosine_warmup, onecycle
    step_size: 10         # for step scheduler
    gamma: 0.1            # LR multiplier for step scheduler
    patience: 5           # for plateau scheduler
    min_lr: 1e-06         # minimum learning rate
    max_lr: 0.01          # for onecycle scheduler

  # Evaluation Metric
  eval_metric: auc        # Options: accuracy, auc, f1, recall, precision
\end{lstlisting}
\section{Validation and Performance}  
\label{sec:5}

To assess the efficiency of \coll for code generation, we conduct a systematic validation study using a wide range of final state particles.
All validation results presented in this section are obtained using the \texttt{meta-llama/Llama-3.3-70B-Instruct} model.
Five benchmark analyses are considered, each benchmark targeting a different physics process and testing a complementary set of selection criteria, kinematic observables and reconstruction methods.
The corresponding test event files and user input files are distributed with the \coll package in the \texttt{data/} and \texttt{examples\_user\_input/} directories, enabling independent verification of all results and providing a broad set of reference distributions for user studies.
For each benchmark, Monte Carlo event samples are generated using \texttt{MadGraph5\_aMC@NLO}~\cite{Alwall:2014hca,Frederix:2018nkq} for parton level event generation, \texttt{Pythia8}~\cite{Bierlich:2022pfr} for parton showering and hadronisation, and \texttt{Delphes3}~\cite{deFavereau:2013fsa} for fast detector simulation with the default CMS detector card. The  generated test files are:

\begin{itemize}
    \item \texttt{signal\_1.lhco}: $pp \to W^+W^-$, with $W^+ \to \ell^+\nu$ and $W^- \to jj$
    \item \texttt{signal\_2.lhco}: $pp \to t\bar{t}$, with $t \to bW^+ \to b\ell^+\nu$ and $\bar{t} \to \bar{b}W^- \to \bar{b}jj$
    \item \texttt{signal\_3.lhco}: $pp \to H \to \gamma\gamma$ 
    \item \texttt{signal\_4.lhco}: $pp \to WZ$, with $W \to \ell\nu$ and $Z \to \ell^+\ell^-$
    \item \texttt{signal\_5.lhco}: $pp \to Hjj$, with $H \to \tau^+\tau^-$
\end{itemize}

\noindent
Each sample contains approximately $\mathcal{O}(10^4)$ events. This size is sufficient to validate the selection logic and to perform visual checks of the main kinematic distributions.

\subsection{Benchmark analyses}
\label{subsec:benchmarks}

The five benchmark analyses are designed to test complementary aspects of the \coll code generation framework.
Table~\ref{tab:benchmark_summary} summarises the considered physics processes, the main selection requirements and the number of selection cuts and validation plots for each example.
These  examples exhibit the following code generation capabilities: lepton, jet, photon, and tau identification and kinematic selection; $b$-jet tagging and vetoing; missing transverse energy requirements; invariant and transverse mass reconstruction; opposite-sign same-flavour lepton pairing; angular separation ($\Delta R$, $\Delta\phi$, $\Delta\eta$) calculations; isolation and overlap removal logic; derived kinematic variables such as $H_T$, $p_T/m_{\gamma\gamma}$ ratios, and the Collins--Soper angle $|\cos\theta^*|$; as well as cutflow reporting and CSV export for subsequent deep learning analyses.

\begin{table}[thbp]
\centering
\caption{Summary of the five benchmark analyses used for validation. For each example, the table lists the physics process, the primary final state objects, the total number of specified selection cuts ($N_{\text{cuts}}$), and the number of requested validation plots ($N_{\text{plots}}$).}
\label{tab:benchmark_summary}
\renewcommand{\arraystretch}{1.25}
\small
\begin{tabular}{@{}c l l c c@{}}
\toprule
\textbf{Example} & \textbf{Process} & \textbf{Final state} & $N_{\text{cuts}}$ & $N_{\text{plots}}$ \\
\midrule
1 & $pp \to W^+W^-$ (semi-leptonic) & $1\ell + 2j + E_T^{\text{miss}}$ & 9 & 10 \\
2 & $pp \to t\bar{t}$ (semi-leptonic)  & $1\ell + \geq 4j\,(2b) + E_T^{\text{miss}}$ & 7 & 10 \\
3 & $pp \to H \to \gamma\gamma$        & $2\gamma$ & 8 & 9 \\
4 & $pp \to WZ$ (trilepton)            & $3\ell + E_T^{\text{miss}}$ & 8 & 10 \\
5 & $pp \to Hjj,\; H \to \tau^+\tau^-$ (VBF) & $2\tau + 2j$ & 9 & 11 \\
\bottomrule
\end{tabular}
\end{table}

To describe the validation of the code generation pipeline, we  take Example~1, 
($p p \rightarrow W^+ W^-$, $W^+ \rightarrow  l + \nu$, $W^- \rightarrow j j$)
as the benchmark for the reproducibility study. The {\ttfamily template/user\_input\_1.txt}  aimed for the analysis of this process, 
\begin{lstlisting}[style=prompt, %title=user\_input\_1.tx, %numbers=none,basicstyle=\ttfamily\small, frame=single, %backgroundcolor=\color{gray!10}
]
[SELECTION CUTS]
- Select electrons with pT > 25 GeV and |eta| < 2.5
- Select muons with  p_T > 20 GeV and | eta |  < 2.4
- Require exactly 1 lepton (electron or muon)
- Require at least 2 jets with pT > 30 GeV and |eta < 2.5
-  Select MET >30 GeV
- Veto events with b-tagged jets 
- Select transverse mass of lepton + MET between 40 and 120 GeV 
(W leptonic candidate)
- Select invariant mass of the two leading jets between 60 and 100 GeV 
(W hadronic candidate)
- Select delta R between the two leading jets < 3.0
....
....
....
[OUTPUT STRUCTURE]
- Print cutflow showing the number of events after each selection
-Save the plots in png format
....
\end{lstlisting}

The analysis is repeated several times under identical conditions, using the same user input specification, the same LHCO data file (\texttt{data/signal\_1.lhco}), the same maximum number of code fixing trials (5), and the same LLM configuration (\texttt{Llama-3.3-70B-Instruct}), to observe the generated analysis code. 
Since the generation process employs strictly deterministic decoding, a high level of consistency is expected across repeated runs executed on the same hardware and within the same software environment. However, variations may arise from non-deterministic code fixing process.  Therefore, the instruction of printing cutflow in {\ttfamily[OUTPUT STRUCTURE]} is useful to check the generated analysis codes.

\begin{table}[h]
\begin{center}
\caption{Cutflows generated from different instance using \texttt{user\_input\_1.txt} for Events \texttt{data/signal\_1.lhco}.}
\label{app:numerical}
\begin{tabular}{|l|rrr|}
\hline
 & A & B & C \cr
\hline
Initial events: &30000& 30000 & 30000 \cr
Events after lepton selection: &13722 & 13722& 15866 \cr
Events after jet selection:& 5729 & 5729 &23528\cr
Events after MET selection: &3935 & 3935 &21031 \cr
Events after b-jet veto:& 3431&3431 &27328\cr
Events after W leptonic selection:& 2943&2943 &10823\cr
Events after W hadronic selection: &{\it 1501} &{\it 1553} & 11053\cr
Events after delta R selection: &{\it 1334}& {\it 1381} & 17052\cr
Events after all selection &-- &-- &{\it 1381} \cr 
\hline 
\end{tabular}
\end{center}
\end{table}

The numerical output of several instance are shown in Table \ref{app:numerical}. 
In Case C, the code output reports the number of events after each cut,  and the number of events remaining after all selection cuts is shown at the end. This behavior is reasonable when the definition of the ``cutflow'' is not precisely specified. 

In Cases A and B, cuts are applied sequentially, following the order given in the user input\footnote{ The number of events after "lepton selection"  defined differently  between C and A(B), causing mismatch of the numbers. }. 
The numerical difference between A and B begins at the stage of the W hadronic selection.
Among the ten independent analysis of sequential cut application, Case A appears in only one instance, while the remaining nine correspond to Case B. 

Both A and B are correct for the given prompt. The prompt {\ttfamily user\_input\_1.txt} contains ambiguity at the definition of ``the two leading jets'' used in the jet invariant mass calculation, and this ambiguity is the primary source of the differing behaviors. 
 In Script~A, the code first selects jets that satisfy the basic jet cuts on $\eta$ and $p_T$, and then computes the
invariant mass of the two leading jets passing the basic jet cuts.
\begin{lstlisting}[language=Python
%,showstringspaces=false,numbers=none
]
       # Jet selection
        jets = get_jets(event)
        selected_jets = [j for j in jets if j['pt'] > 30 and abs(j['eta']) < 2.5]
        if len(selected_jets) < 2:
            continue
        events_after_jet_selection += 1
        ....
        ....
        # W hadronic selection
        leading_jet = get_leading(selected_jets)
        subleading_jet = get_subleading(selected_jets)
        m_jj_event = invariant_mass([leading_jet, subleading_jet])
        if m_jj_event < 60 or m_jj_event > 100:
            continue
        events_after_w_hadronic_selection += 1
\end{lstlisting}
On the other hand, the generated scripts corresponding to Case B select the events having more than two jets that satisfy the $p_T$ and $\eta$ condition, then calculate the invariant mass of two leading jet without requiring condition on $\eta$:
\begin{lstlisting}[language=Python, %showstringspaces=false,numbers=none
]
def dijet_mass(event):
    """Compute invariant mass of two leading jets."""
    jets = sort_by_pt(get_jets(event))
    if len(jets) < 2:
        return 0.0
    
    # Sum 4-momenta (NOT difference!)
    E1, px1, py1, pz1 = four_momentum(jets[0])
    E2, px2, py2, pz2 = four_momentum(jets[1])
    
    E_tot = E1 + E2
    px_tot = px1 + px2
    py_tot = py1 + py2
    pz_tot = pz1 + pz2
    
    M_sq = E_tot**2 - px_tot**2 - py_tot**2 - pz_tot**2
    return math.sqrt(M_sq) if M_sq > 0 else 0.0
   ....
   ....
   # Require at least 2 jets
    selected_events = [event for event in selected_events 
    if len([j for j in get_jets(event) if j['pt'] > 30 and 
    abs(j['eta']) < 2.5]) >= 2]
    print(f"Events after jet selection: {len(selected_events)}")
   ....
   ....
    # Select invariant mass of the two leading jets between 60 and 100 GeV
    selected_events = [event for event in selected_events if 60 < dijet_mass(event) < 100]
    print(f"Events after W hadronic selection: {len(selected_events)}")
   ....
   ....
\end{lstlisting}

The differences among the generated codes are often subtle, and analyzing them manually is inefficient. 
Since the generated scripts in both cases consist of approximately 300 lines of code, we recommend that users employ an external LLM to assist in reviewing and validating the generated analysis, as discussed in appendix~\ref{app:1}.

\begin{figure}[!h]
\centering
\includegraphics[width=\textwidth]{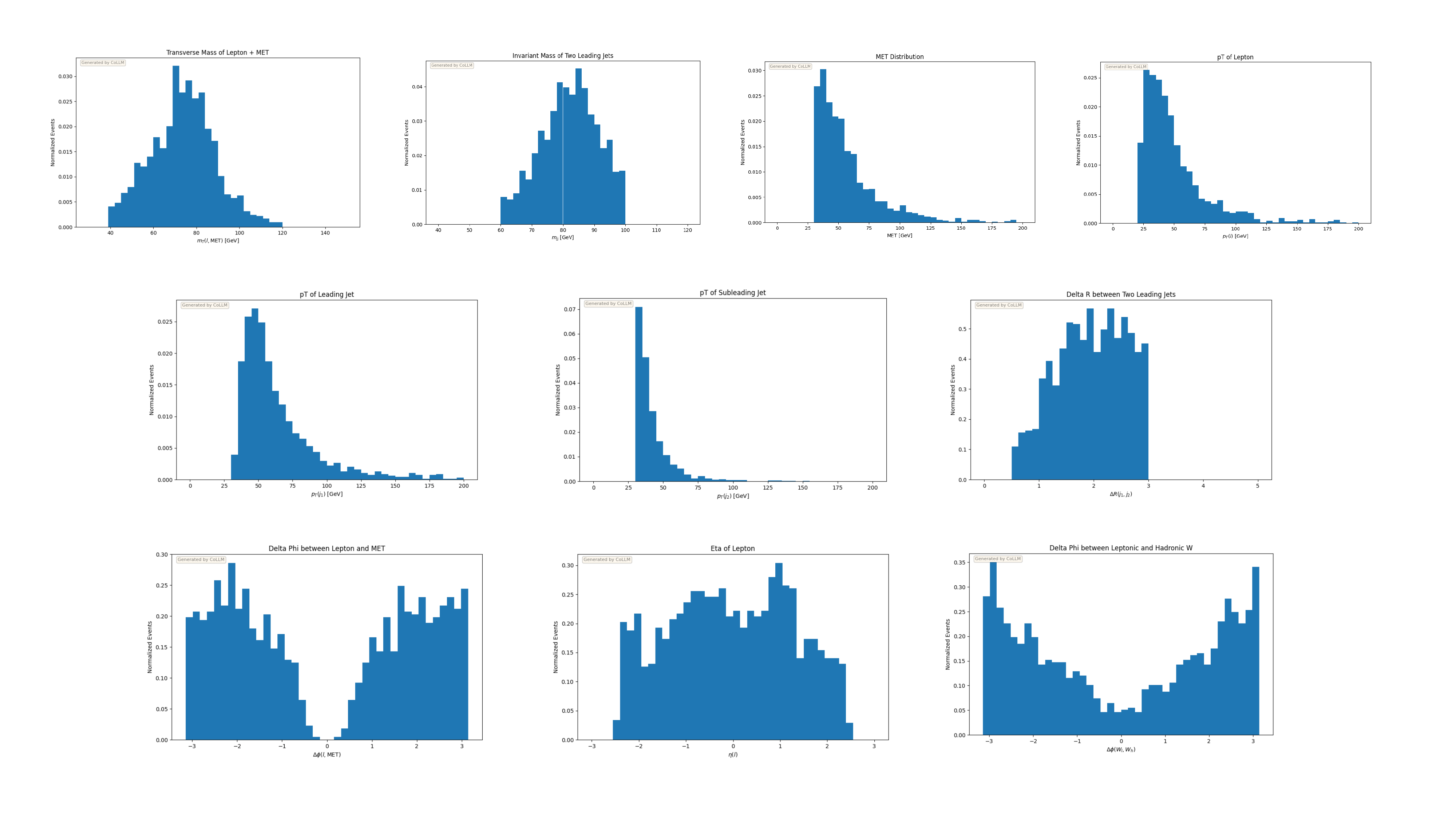} 
\caption{Validation histograms generated by \coll for Example 1.}
\label{fig:ex4_plots}
\end{figure}

The ten validation histograms produced by Example 1, case B are shown in Figure~\ref{fig:ex4_plots}. These distributions serve as a physics level sanity check of the generated analysis code.

\subsection{Comparison with LLM rapid code generation}
\label{subsec:comparison_llms}

Generic LLMs have demonstrated strong capabilities in code generation and are increasingly used in high energy physics searches, documentation, and exploratory studies. However, when applied to full collider analysis workflows, they face fundamental limitations related to automation, physical correctness and reproducibility.
A central limitation is the absence of an integrated analysis pipeline. Generic LLMs operate as stateless prompts and cannot natively coordinate sequential stages such as data parsing, object reconstruction, event selection, validation, and execution. These steps must therefore be connected manually, increasing development time and the risk of inconsistencies. Moreover, Generic LLMs lack embedded domain knowledge and do not explicitly encode standard high-energy physics conventions. As a result, generated code may contain subtle conceptual errors, including incorrect kinematic expressions, inconsistent units, or confusion between detector-level and truth-level quantities.
Reproducibility also presents a challenge, as LLM outputs are inherently non-deterministic and may vary across sessions, prompts, or model versions.

\begin{table}[!h]
\centering
\caption{Comparison of \coll with generic LLMs for collider analysis.}
\label{tab:comparison}
\renewcommand{\arraystretch}{1.15}
\small
\begin{tabularx}{\textwidth}{@{}l >{\raggedright\arraybackslash}X >{\raggedright\arraybackslash}X@{}}
\toprule
\textbf{Feature} & \textbf{\coll} & \textbf{Generic LLMs} \\
\midrule
End-to-End Pipeline & 
 Integrated analysis workflow & 
 Manual and fragmented \\
\addlinespace

Physics Awareness & 
 Built-in HEP conventions and observables & 
 Generic, error-prone \\
\addlinespace

Code Execution and Validation & 
 Automatic execution and self-correction & 
 No execution or validation \\
\addlinespace

LHCO and Data Handling & 
 Native file parsing and batch processing & 
No direct data support \\
\addlinespace

Local Deployment & 
 local GPU accelerated and cloud based & 
 Cloud based services \\
\addlinespace

Reproducibility & 
 Deterministic and version controlled & 
 Non-deterministic outputs \\
\addlinespace

Deep Learning Integration & 
 Integrated training and configuration & 
 No training support \\
\bottomrule
\end{tabularx}
\end{table}

Table~\ref{tab:comparison} summarizes these differences between generic LLMs and \coll. While the former primarily function as conversational assistants, \coll is designed as a complete analysis framework.
A key advantage of \coll is its embedded domain expertise, encoded in the system prompt through detailed knowledge of standard formats and physics observables. Reliability is further enhanced by the PyFixer module, which implements an automated generate--validate--correct loop for error detection and correction.

In summary, while generic LLMs remain useful for auxiliary tasks, they do not satisfy the core requirements of collider analyses. \coll addresses these limitations through its specialized architecture and validation pipeline, enabling reliable and reproducible physics analyses.

\section{Discussion and conclusion }  
\label{sec:6}
In this work, we presented \coll, a public framework that connects plain language analysis specifications to automated deep learning pipelines for collider event data. The framework integrates physics aware LLM code generation with MLP, GNN and Transformer architectures. Using five benchmark processes covering different final states and kinematic regimes, we demonstrated that \coll produces consistent and executable analysis code with physically meaningful outputs. Deterministic decoding, together with the \texttt{PyFixer} error correction mechanism, ensures a high level of reproducibility across repeated runs.

Despite these capabilities, the current implementation has several limitations. Code generation is restricted to the LHCO data format and does not yet support ROOT  structure, commonly used in experimental analyses. Extending \coll to this format will require adapting both the system prompt and the code templates to more complex data representations.

Plain language ambiguity remains an inherent challenge. As shown in Section~\ref{sec:5} and Appendix~\ref{app:1},  physically ambiguous interpretations of the same prompt may lead to different implementations and numerical results. Although the structured input format mitigates this effect, it does not eliminate it entirely. Future versions could incorporate interactive clarification procedures and automated validation against reference implementations. 


The current GNN module employs fully connected graphs with trivial edge features and does not exploit the internal correlation between particles. Incorporating adaptive graph constructions and physically informed edge features, such as angular separations and invariant masses, is expected to improve performance for complex final states. 

Model hyperparameters are currently defined by the user, with no automated optimisation. As a result, classifier performance may be suboptimal for non-standard analyses. Integrating tuning methods would reduce user intervention and improve robustness. 

In summary, \coll demonstrates that domain-aware LLM-based code generation combined with automated deep learning pipelines can substantially reduce the technical overhead of collider analyses while preserving reproducibility and physical reliability. The framework is publicly available and is designed to evolve alongside advances in deep learning and language models for collider physics.

\section*{Acknowledgments}
 This work is funded by grant number 22H05113, ``Foundation of Machine Learning Physics'', Grant in Aid for Transformative Research Areas and 22K03626, Grant-in-Aid for Scientific Research (C). WE is funded by the ErUM-WAVE project 05D2022, “ErUM-Wave: Antizipation 3-dimensionaler Wellenfelder”, which is supported by the German Federal Ministry of Education and Research (BMBF).

\appendix
\section{Using an external LLM for code validation }
\label{app:1}
When expressions in a \texttt{user\_input} file are ambiguous, the generated code may behave differently from the user’s intention. Although the package provides several methods for testing the generated code, it is often useful to prompt an external AI chatbot to analyze the code in greater depth. Here, we present an example in which ChatGPT-5.2 Pro is introduced as an external code reviewer, while the code itself is generated by our best-performing LLM model, \texttt{Llama-3.3-70B-Instruct}. Given an ambiguous prompt, the LLM generates codes that produce different numerical outputs, and the reviewer, ChatGPT, identifies the underlying differences between the generated codes.

Here, we study the generated codes discussed in Section 5.2.  The differences among the generated python codes A and B are subtle, and manual analysis is inefficient. Generic LLM can compare two generated codes, identify their differences, and guide improvements to the prompts without requiring access to other system files. Let ask ChatGPT;
\begin{lstlisting}[ style=prompt, %basicstyle=\ttfamily\small, breaklines=true,
%frame=single, 
%backgroundcolor=\color{gray!10},showstringspaces=false, numbers=none
]
Please analyze these two codes. The output counts for "Events after 
W hadronic selection:" and "Events after delta R selection:"
are different. What is the reason for this difference?
\end{lstlisting}
Then one would get follwing; (output may vary with personal settings) 
\begin{tcolorbox}[breakable,
  colback=gray!10,
  colframe=gray!60,
  boxrule=0.5pt,
  arc=2pt,
  left=6pt,
  right=6pt,
  top=6pt,
  bottom=6pt
]
\textbf{Reason for the difference between the event counts}

\medskip
.... 

\noindent 
....
\medskip 

\textbf{2.\ Different jet collections are used in the two codes}

This is the dominant source of the discrepancy \emph{between the two implementations}:

\begin{itemize}
  \item In the first code, jets are explicitly filtered,
  \[
  p_T > 30~\mathrm{GeV}, \qquad |\eta| < 2.5,
  \]
  and both \(m_{jj}\) and \(\Delta R_{jj}\) are computed using the two leading jets from this \emph{selected} jet list.

  \item In the second code, the event is first required to have at least two jets passing the above kinematic criteria, but the functions computing
  \(m_{jj}\) and \(\Delta R_{jj}\) use the \emph{full jet collection} (sorted by \(p_T\)) without reapplying the selection.

\medskip 
....

\noindent
....
\end{itemize}
\end{tcolorbox}
\noindent 
Alternatively, one may ask to provide a more precise list of the cut conditions implemented in the code. Below, we show an example output, 
\begin{tcolorbox}[breakable,
  colback=gray!10,
  colframe=gray!60,
  boxrule=0.5pt,
  arc=2pt,
  left=6pt,
  right=6pt,
  top=6pt,
  bottom=6pt
]

The script prints a cutflow summary consisting of eight lines, each reporting the cumulative number of events surviving a given selection step:

\begin{enumerate}
  \item \texttt{Initial events: $N_0$} \\
  Total number of events read from the LHCO input file.

  \item \texttt{Events after lepton selection: $N_1$} \\
  Events containing exactly one selected lepton, defined as:
  \begin{itemize}
    \item Electrons: $\texttt{type}=1$, $p_T > 25~\mathrm{GeV}$, $|\eta| < 2.5$
    \item Muons: $\texttt{type}=2$, $p_T > 20~\mathrm{GeV}$, $|\eta| < 2.4$
  \end{itemize}
  The requirement is:
  $
    N_\ell = N_e + N_\mu = 1 .
  $

  \item \texttt{Events after jet selection: $N_2$} \\
  Events passing the lepton selection and containing at least two jets satisfying:
  $
    \texttt{type}=4,\quad p_T > 30~\mathrm{GeV},\quad |\eta| < 2.5 .
  $

  \item \texttt{Events after MET selection: $N_3$} \\
  Events passing previous steps and satisfying:
  $
    E_T^{\text{miss}} > 30~\mathrm{GeV},
  $
  where MET corresponds to the object with $\texttt{type}=6$.

  \item \texttt{Events after b-jet veto: $N_4$} \\
  Events with no b-tagged jets, where a b-jet is defined as:
  $
    \texttt{type}=4 \ \text{and} \ \texttt{btag}=1.0 .
  $

  \item \texttt{Events after W leptonic selection: $N_5$} \\
  Events satisfying a transverse mass requirement for the lepton--MET system:
  $
    m_T = \sqrt{2 p_T^\ell E_T^{\text{miss}}
      \left(1 - \cos \Delta\phi(\ell, E_T^{\text{miss}})\right)},
  $
  with
  $
    40~\mathrm{GeV} < m_T < 120~\mathrm{GeV}.
  $

  \item \texttt{Events after W hadronic selection: $N_6$} \\
  Events whose dijet invariant mass satisfies:
  $
    60~\mathrm{GeV} < m_{jj} < 100~\mathrm{GeV}.
  $
  The invariant mass $m_{jj}$ is computed using the two highest-$p_T$ jets
  in the \emph{entire event}, without applying additional $p_T$ or $\eta$ cuts.
  Jet four-momenta are constructed as:
  $
    \begin{aligned}
      p_x &= p_T \cos\phi, \\
      p_y &= p_T \sin\phi, \\
      p_z &= p_T \sinh\eta, \\
      E   &= \sqrt{p_x^2 + p_y^2 + p_z^2 + m^2}.
    \end{aligned}
  $

  \item \texttt{Events after delta R selection: $N_7$} \\
  Events satisfying an angular separation requirement between the same two jets:
  $
    \Delta R_{jj} = \sqrt{(\Delta\eta)^2 + (\Delta\phi)^2} < 3.0,
  $
  where $\Delta\phi$ is wrapped into the interval $[-\pi, \pi]$.
 
\end{enumerate}
\end{tcolorbox}
\noindent 
Remarkablly,  ChatGPT explicitly lists the definitions of parameters,such as $M_T$, together with the corresponding functions, so that users do not
need to refer back to the script to locate their definitions.

As illustrated above, the use of ChatGPT for detailed code inspection and for summarizing generated outputs helps to accurately extract the semantics of diverse
codes produced from naturally ambiguous prompts. This, in turn, assists users in refining their prompts and provides researchers with insights into how their
analysis workflows are designed and interpreted. 
\bibliographystyle{JHEP}
\bibliography{biblo}

@article{Larkoski:2017jix,
    author = "Larkoski, Andrew J. and Moult, Ian and Nachman, Benjamin",
    title = "{Jet Substructure at the Large Hadron Collider: A Review of Recent Advances in Theory and Machine Learning}",
    journal = "Phys. Rept.",
    volume = "841",
    pages = "1--63",
    year = "2020",
    doi = "10.1016/j.physrep.2019.11.001",
    eprint = "1709.04464",
    archivePrefix = "arXiv",
    primaryClass = "hep-ph"
}

@article{Guest:2018yhq,
    author = "Guest, Dan and Cranmer, Kyle and Whiteson, Daniel",
    title = "{Deep Learning and its Application to LHC Physics}",
    journal = "Ann. Rev. Nucl. Part. Sci.",
    volume = "68",
    pages = "161--181",
    year = "2018",
    doi = "10.1146/annurev-nucl-101917-021019",
    eprint = "1806.11484",
    archivePrefix = "arXiv",
    primaryClass = "hep-ex"
}

@article{Radovic:2018dip,
    author = "Radovic, Alexander and others",
    title = "{Machine learning at the energy and intensity frontiers of particle physics}",
    journal = "Nature",
    volume = "560",
    pages = "41--48",
    year = "2018",
    doi = "10.1038/s41586-018-0361-2"
}

@article{Bourilkov:2019yoi,
    author = "Bourilkov, Dimitri",
    title = "{Machine and Deep Learning Applications in Particle Physics}",
    journal = "Int. J. Mod. Phys. A",
    volume = "34",
    pages = "1930019",
    year = "2019",
    doi = "10.1142/S0217751X19300199",
    eprint = "1912.08245",
    archivePrefix = "arXiv",
    primaryClass = "physics.data-an"
}

@article{Feickert:2021ajf,
    author = "Feickert, Matthew and Nachman, Benjamin",
    title = "{A Living Review of Machine Learning for Particle Physics}",
    year = "2021",
    eprint = "2102.02770",
    archivePrefix = "arXiv",
    primaryClass = "hep-ph"
}

@article{brown2020language,
    author = "Brown, Tom B. and others",
    title = "{Language Models are Few-Shot Learners}",
    journal = "Adv. Neural Inf. Process. Syst.",
    volume = "33",
    pages = "1877--1901",
    year = "2020",
    eprint = "2005.14165",
    archivePrefix = "arXiv",
    primaryClass = "cs.CL"
}

@article{openai2023gpt4,
    author = "{OpenAI}",
    title = "{GPT-4 Technical Report}",
    year = "2023",
    eprint = "2303.08774",
    archivePrefix = "arXiv",
    primaryClass = "cs.CL"
}

@article{touvron2023llama,
    author = "Touvron, Hugo and others",
    title = "{LLaMA: Open and Efficient Foundation Language Models}",
    year = "2023",
    eprint = "2302.13971",
    archivePrefix = "arXiv",
    primaryClass = "cs.CL"
}

@article{llama3,
    author = "Grattafiori, Aaron and others",
    title = "{The Llama 3 Herd of Models}",
    year = "2024",
    eprint = "2407.21783",
    archivePrefix = "arXiv",
    primaryClass = "cs.AI"
}

@article{qwen2.5,
    author = "Yang, An and others",
    title = "{Qwen2.5 Technical Report}",
    year = "2025",
    eprint = "2412.15115",
    archivePrefix = "arXiv",
    primaryClass = "cs.CL"
}

@article{deepseek2025,
    author = "Guo, Daya and others",
    title = "{DeepSeek-R1: Incentivizing Reasoning Capability in LLMs via Reinforcement Learning}",
    year = "2025",
    eprint = "2501.12948",
    archivePrefix = "arXiv",
    primaryClass = "cs.CL"
}

@article{Simons:2024astro,
    author = "Simons, Arno",
    title = "{Astro-HEP-BERT: A bidirectional language model for studying the meanings of astrophysics and high energy physics concepts}",
    year = "2024",
    eprint = "2407.00007",
    archivePrefix = "arXiv",
    primaryClass = "cs.CL"
}

@article{Richmond:2025feyntune,
    author = "Richmond, Paul and Agarwal, Prarit and Chowdhury, Borun and Niarchos, Vasilis and Papageorgakis, Constantinos",
    title = "{FeynTune: Large Language Models for High-Energy Theory}",
    year = "2025",
    eprint = "2502.08625",
    archivePrefix = "arXiv",
    primaryClass = "hep-th"
}

@article{Fanelli:2024eic,
    author = "Fanelli, Cristiano and others",
    title = "{Physics Event Classification Using Large Language Models}",
    year = "2024",
    eprint = "2404.05752",
    archivePrefix = "arXiv",
    primaryClass = "hep-ex"
}

@article{Ndum:2024autofluka,
    author = "Ndum, Emmanuel and others",
    title = "{AutoFLUKA: A Large Language Model Based Framework for Automating Monte Carlo Simulations in Particle Physics}",
    year = "2024",
    eprint = "2407.16681",
    archivePrefix = "arXiv",
    primaryClass = "physics.comp-ph"
}

@article{Diefenbacher:2025zzn,
    author = "Diefenbacher, Sascha and others",
    title = "{Agents of Discovery}",
    year = "2025",
    eprint = "2504.16736",
    archivePrefix = "arXiv",
    primaryClass = "hep-ph"
}

@article{HWresearch:2025llm4hep,
    author = "Wang, Hao and others",
    title = "{Automating High Energy Physics Data Analysis with LLM-Powered Agents}",
    year = "2025",
    eprint = "2512.07785",
    archivePrefix = "arXiv",
    primaryClass = "hep-ph"
}

@article{Golling:2024abg,
    author = "Golling, Tobias and others",
    title = "{Large Physics Models: Towards a collaborative approach with Large Language Models and Foundation Models}",
    journal = "Eur. Phys. J. C",
    volume = "85",
    pages = "856",
    year = "2025",
    doi = "10.1140/epjc/s10052-025-14707-8",
    eprint = "2501.05382",
    archivePrefix = "arXiv",
    primaryClass = "hep-ph"
}

@misc{lhco2006,
    author = "Thaler, Jesse",
    title = "{LHC Olympics wiki}",
    howpublished = "\url{https://lhco.hepforge.org}",
    year = "2006",
    note = "Accessed: 2025-01-01"
}

@article{qwen25,
  title        = {Qwen2.5: A Suite of Large Language Models},
  author       = {Qwen Team},
  eprint = "2409.12186",
  archivePrefix = "arXiv",
  year         = {2024}
}

@article{qwen3,
  title={Qwen3 technical report},
  author={Yang, An and Li, Anfeng and Yang, Baosong and Zhang, Beichen and Hui, Binyuan and Zheng, Bo and Yu, Bowen and Gao, Chang and Huang, Chengen and Lv, Chenxu and others},
  eprint = "2505.09388",
  archivePrefix = "arXiv",
    year={2025}
}

@article{deepseekr1,
  title        = {DeepSeek-R1: Incentivizing Reasoning Capability in Large Language Models via Reinforcement Learning},
  author       = {DeepSeek AI},
 eprint = "2501.12948",
  archivePrefix = "arXiv",
  year         = {2025}
}

@article{qwq,
  title={Reasoning language models: A blueprint},
  author={Besta, Maciej and Barth, Julia and Schreiber, Eric and Kubicek, Ales and Catarino, Afonso and Gerstenberger, Robert and Nyczyk, Piotr and Iff, Patrick and Li, Yueling and Houliston, Sam and others},
 eprint = "2501.11223",
  archivePrefix = "arXiv",
  year={2025}
}

@article{ge2025survey,
  title        = {A Survey of Vibe Coding with Large Language Models},
  author       = {Yuyao Ge and Lingrui Mei and Zenghao Duan and Tianhao Li and Yujia Zheng and Yiwei Wang and Lexin Wang and Jiayu Yao and Tianyu Liu and Yujun Cai and Baolong Bi and Fangda Guo and Jiafeng Guo and Shenghua Liu and Xueqi Cheng},
 eprint = "2510.12399",
  archivePrefix = "arXiv",
  year         = {2025}
}

@inproceedings{Gendreau-Distler:2025fsj,
    author = "Gendreau-Distler, Eli and Ho, Joshua and Kim, Dongwon and Le Pottier, Luc Tomas and Wang, Haichen and Yang, Chengxi",
    title = "{Automating High Energy Physics Data Analysis with LLM-Powered Agents}",
    booktitle = "{39th Annual Conference on Neural Information Processing Systems}: {Includes Machine Learning and the Physical Sciences (ML4PS)}",
    eprint = "2512.07785",
    archivePrefix = "arXiv",
    primaryClass = "physics.data-an",
    month = "12",
    year = "2025"
}

@article{Plehn:2026gxv,
    author = "Plehn, Tilman and Schiller, Daniel and Schmal, Nikita",
    title = "{MadAgents}",
    eprint = "2601.21015",
    archivePrefix = "arXiv",
    primaryClass = "hep-ph",
    month = "1",
    year = "2026"
}

@article{Alwall:2014hca,
    author = "Alwall, J. and Frederix, R. and Frixione, S. and Hirschi, V. and Maltoni, F. and Mattelaer, O. and Shao, H. -S. and Stelzer, T. and Torrielli, P. and Zaro, M.",
    title = "{The automated computation of tree-level and next-to-leading order differential cross sections, and their matching to parton shower simulations}",
    eprint = "1405.0301",
    archivePrefix = "arXiv",
    primaryClass = "hep-ph",
    reportNumber = "CERN-PH-TH-2014-064, CP3-14-18, LPN14-066, MCNET-14-09, ZU-TH-14-14",
    doi = "10.1007/JHEP07(2014)079",
    journal = "JHEP",
    volume = "07",
    pages = "079",
    year = "2014"
}

@article{Frederix:2018nkq,
    author = "Frederix, R. and Frixione, S. and Hirschi, V. and Pagani, D. and Shao, H. -S. and Zaro, M.",
    title = "{The automation of next-to-leading order electroweak calculations}",
    eprint = "1804.10017",
    archivePrefix = "arXiv",
    primaryClass = "hep-ph",
    reportNumber = "Nikhef/2018-015, TUM-HEP-1138/18, NIKHEF-2018-015, TUM-HEP-1138-18",
    doi = "10.1007/JHEP11(2021)085",
    journal = "JHEP",
    volume = "07",
    pages = "185",
    year = "2018",
    note = "[Erratum: JHEP 11, 085 (2021)]"
}

@article{Bierlich:2022pfr,
    author = "Bierlich, Christian and others",
    title = "{A comprehensive guide to the physics and usage of PYTHIA 8.3}",
    eprint = "2203.11601",
    archivePrefix = "arXiv",
    primaryClass = "hep-ph",
    reportNumber = "LU-TP 22-16, MCNET-22-04, FERMILAB-PUB-22-227-SCD",
    doi = "10.21468/SciPostPhysCodeb.8",
    journal = "SciPost Phys. Codeb.",
    volume = "2022",
    pages = "8",
    year = "2022"
}

@article{deFavereau:2013fsa,
    author = "de Favereau, J. and Delaere, C. and Demin, P. and Giammanco, A. and Lema{\^\i}tre, V. and Mertens, A. and Selvaggi, M.",
    collaboration = "DELPHES 3",
    title = "{DELPHES 3, A modular framework for fast simulation of a generic collider experiment}",
    eprint = "1307.6346",
    archivePrefix = "arXiv",
    primaryClass = "hep-ex",
    doi = "10.1007/JHEP02(2014)057",
    journal = "JHEP",
    volume = "02",
    pages = "057",
    year = "2014"
}

@article{Fuks:2025qgh,
    author = "Fuks, Benjamin and Garg, Sumit K. and Hammad, A. and Jueid, Adil",
    title = "{Deep learning approaches to top FCNC couplings to photons at the LHC}",
    eprint = "2507.17807",
    archivePrefix = "arXiv",
    primaryClass = "hep-ph",
    reportNumber = "CTPU-PTC-25-16",
    month = "7",
    year = "2025"
}

@inproceedings{he2015delving,
  title={Delving deep into rectifiers: Surpassing human-level performance on imagenet classification},
  author={He, Kaiming and Zhang, Xiangyu and Ren, Shaoqing and Sun, Jian},
  booktitle={Proceedings of the IEEE international conference on computer vision},
  pages={1026--1034},
  year={2015}
}

@article{clevert2015fast,
  title={Fast and accurate deep network learning by exponential linear units (elus)},
  author={Clevert, Djork-Arn{\'e} and Unterthiner, Thomas and Hochreiter, Sepp},
   eprint = "1511.07289",
    archivePrefix = "arXiv",
  
  volume={4},
  number={5},
  pages={11},
  year={2015}
}

@article{klambauer2017self,
  title={Self-normalizing neural networks},
  author={Klambauer, G{\"u}nter and Unterthiner, Thomas and Mayr, Andreas and Hochreiter, Sepp},
 
 journal={Advances in neural information processing systems},
  volume={30},
  year={2017}
}

@article{hendrycks2016gaussian,
  title={Gaussian Error Linear Units (Gelus)},
  author={Hendrycks, D},
    eprint = "1606.08415",
    archivePrefix = "arXiv",
  year={2016}
}

@article{Qu:2019gqs,
    author = "Qu, Huilin and Gouskos, Loukas",
    title = "{ParticleNet: Jet Tagging via Particle Clouds}",
    eprint = "1902.08570",
    archivePrefix = "arXiv",
    primaryClass = "hep-ph",
    doi = "10.1103/PhysRevD.101.056019",
    journal = "Phys. Rev. D",
    volume = "101",
    number = "5",
    pages = "056019",
    year = "2020"
}

@article{Komiske:2018cqr,
    author = "Komiske, Patrick T. and Metodiev, Eric M. and Thaler, Jesse",
    title = "{Energy Flow Networks: Deep Sets for Particle Jets}",
    eprint = "1810.05165",
    archivePrefix = "arXiv",
    primaryClass = "hep-ph",
    reportNumber = "MIT-CTP 5064",
    doi = "10.1007/JHEP01(2019)121",
    journal = "JHEP",
    volume = "01",
    pages = "121",
    year = "2019"
}

@article{Mikuni:2024qsr,
    author = "Mikuni, Vinicius and Nachman, Benjamin",
    title = "{Solving key challenges in collider physics with foundation models}",
    eprint = "2404.16091",
    archivePrefix = "arXiv",
    primaryClass = "hep-ph",
    doi = "10.1103/PhysRevD.111.L051504",
    journal = "Phys. Rev. D",
    volume = "111",
    number = "5",
    pages = "L051504",
    year = "2025"
}

@article{Qu:2022mxj,
    author = "Qu, Huilin and Li, Congqiao and Qian, Sitian",
    title = "{Particle Transformer for Jet Tagging}",
    eprint = "2202.03772",
    archivePrefix = "arXiv",
    primaryClass = "hep-ph",
    month = "2",
    year = "2022"
}

@article{Esmail:2025kii,
    author = "Esmail, W. and Hammad, A. and Nojiri, M.",
    title = "{IAFormer: Interaction-Aware Transformer network for collider data analysis}",
    eprint = "2505.03258",
    archivePrefix = "arXiv",
    primaryClass = "hep-ph",
    month = "5",
    year = "2025"
}

@article{Hammad:2024qme,
    author = "Hammad, Ahmed and Nojiri, Mihoko M.",
    title = "{Transformer Networks for Heavy Flavor Jet Tagging}",
    eprint = "2411.11519",
    archivePrefix = "arXiv",
    primaryClass = "hep-ph",
    doi = "10.7566/JPSJ.94.031007",
    journal = "J. Phys. Soc. Jap.",
    volume = "94",
    number = "3",
    pages = "031007",
    year = "2025"
}

@article{Hammad:2023sbd,
    author = "Hammad, A. and Moretti, S. and Nojiri, M.",
    title = "{Multi-scale cross-attention transformer encoder for event classification}",
    eprint = "2401.00452",
    archivePrefix = "arXiv",
    primaryClass = "hep-ph",
    doi = "10.1007/JHEP03(2024)144",
    journal = "JHEP",
    volume = "03",
    pages = "144",
    year = "2024"
}

@article{vaswani2017attention,
  title={Attention is all you need},
  author={Vaswani, Ashish and Shazeer, Noam and Parmar, Niki and Uszkoreit, Jakob and Jones, Llion and Gomez, Aidan N and Kaiser, {\L}ukasz and Polosukhin, Illia},
  journal={Advances in neural information processing systems},
  volume={30},
  year={2017}
}

@article{Esmail:2024jdg,
    author = "Esmail, W. and Hammad, A. and Nojiri, M. and Scherb, Christiane",
    title = "{Testing CP properties of the Higgs boson coupling to {\ensuremath{\tau}} leptons with heterogeneous graphs}",
    eprint = "2409.06132",
    archivePrefix = "arXiv",
    primaryClass = "hep-ph",
    doi = "10.1007/JHEP04(2025)083",
    journal = "JHEP",
    volume = "04",
    pages = "083",
    year = "2025"
}

@article{Esmail:2023axd,
    author = "Esmail, W. and Hammad, A. and Moretti, S.",
    title = "{Sharpening the A {\textrightarrow} Z$^{(*)}$h signature of the Type-II 2HDM at the LHC through advanced Machine Learning}",
    eprint = "2305.13781",
    archivePrefix = "arXiv",
    primaryClass = "hep-ph",
    doi = "10.1007/JHEP11(2023)020",
    journal = "JHEP",
    volume = "11",
    pages = "020",
    year = "2023"
}

@article{wang2018dynamic,
  title={Dynamic Graph CNN for Learning on Point Clouds},
  author={Wang, Yue and Sun, Yongbin and Liu, Ziwei and Sarma, Sanjay E. and Bronstein, Michael M. and Solomon, Justin M.},
  journal={arXiv preprint arXiv:1801.07829},
  year={2018}
}

@inproceedings{velickovic2018graph,
  title={Graph Attention Networks},
  author={Veli{\v{c}}kovi{\'c}, Petar and Cucurull, Guillem and Casanova, Arantxa and Romero, Adriana and Li{\`o}, Pietro and Bengio, Yoshua},
  booktitle={6th International Conference on Learning Representations (ICLR)},
  year={2018},
  url={https://arxiv.org/abs/1710.10903}
}

@inproceedings{Spinner:2024hjm,
    author = "Spinner, Jonas and Bres{\'o}, Victor and de Haan, Pim and Plehn, Tilman and Thaler, Jesse and Brehmer, Johann",
    title = "{Lorentz-Equivariant Geometric Algebra Transformers for High-Energy Physics}",
    booktitle = "{38th conference on Neural Information Processing Systems}",
    eprint = "2405.14806",
    archivePrefix = "arXiv",
    primaryClass = "physics.data-an",
    reportNumber = "MIT-CTP/5723",
    month = "10",
    year = "2024"
}

@article{Brehmer:2024yqw,
    author = "Brehmer, Johann and Bres{\'o}, V{\'\i}ctor and de Haan, Pim and Plehn, Tilman and Qu, Huilin and Spinner, Jonas and Thaler, Jesse",
    title = "{A Lorentz-equivariant transformer for all of the LHC}",
    eprint = "2411.00446",
    archivePrefix = "arXiv",
    primaryClass = "hep-ph",
    reportNumber = "MIT-CTP/5802",
    doi = "10.21468/SciPostPhys.19.4.108",
    journal = "SciPost Phys.",
    volume = "19",
    number = "4",
    pages = "108",
    year = "2025"
}

@article{Wu:2024thh,
    author = "Wu, Yifan and Wang, Kun and Li, Congqiao and Qu, Huilin and Zhu, Jingya",
    title = "{Jet tagging with more-interaction particle transformer*}",
    eprint = "2407.08682",
    archivePrefix = "arXiv",
    primaryClass = "hep-ph",
    doi = "10.1088/1674-1137/ad7f3d",
    journal = "Chin. Phys. C",
    volume = "49",
    number = "1",
    pages = "013110",
    year = "2025"
}

@article{He:2023cfc,
    author = "He, Minxuan and Wang, Daohan",
    title = "{Quark/gluon discrimination and top tagging with dual attention transformer}",
    eprint = "2307.04723",
    archivePrefix = "arXiv",
    primaryClass = "hep-ph",
    doi = "10.1140/epjc/s10052-023-12293-1",
    journal = "Eur. Phys. J. C",
    volume = "83",
    number = "12",
    pages = "1116",
    year = "2023"
}

@inproceedings{kipf2017semi,
    author    = "Kipf, Thomas N. and Welling, Max",
    title     = "Semi-Supervised Classification with Graph Convolutional Networks",
    booktitle = "International Conference on Learning Representations (ICLR)",
    year      = "2017",
    doi       = "10.48550/arXiv.1609.02907"
}

@article{wang2019dynamic,
    author  = "Wang, Yue and Sun, Yongbin and Liu, Ziwei and Sarma, Sanjay E. and Bronstein, Michael M. and Solomon, Justin M.",
    title   = "Dynamic Graph CNN for Learning on Point Clouds",
    journal = "ACM Transactions on Graphics (TOG)",
    volume  = "38",
    number  = "5",
    pages   = "1--12",
    year    = "2019",
    doi     = "10.1145/3326362"
}
\end{document}